\documentclass[aps,prd,onecolumn,superscriptaddress,nofootinbib]{revtex4-2}
\usepackage[utf8]{inputenc}
\usepackage{amsbsy}
\usepackage{amsmath}
\usepackage{amsfonts}
\usepackage{xcolor}
\usepackage{physics}
\usepackage{graphicx}

\begin{document}

\title{Boson mixing and flavor  vacuum in the expanding Universe: a possible candidate for the dark energy}
\author{Antonio Capolupo}
\author{Aniello Quaranta}
\affiliation{Dipartimento di Fisica "E.R. Caianiello" Universit\'a di Salerno,  and INFN - Gruppo Collegato di Salerno, Italy}

\begin{abstract}

We analyze the boson mixing in curved spacetime and  compute the expectation value of the energy-momentum tensor of bosons on the flavor vacuum in spatially flat Friedmann-Lemaître-Robertson-Walker metrics. We show that the energy-momentum
tensor of the flavor vacuum behaves as the effective
energy-momentum tensor of a perfect fluid. Assuming a fixed de Sitter background, we show that the
equation of state  can assume values in the interval $[-1, 1]$, and, in the flat space-time limit has a value $-1$, which is the one of the dark energy. The results here presented show that vacuum of mixed bosons like neutrino super-partners, can represent a possible component of the dark energy of the Universe.

\end{abstract}

\maketitle

\section{Introduction}

One of the most important discoveries of modern cosmology is represented by the observation of the accelerated expansion of the universe \cite{CMBR,Spergel,LSS,Szalay,DE1,Peebles,SNeIa}. In the context of general relativity, classical sources of matter generate only positive pressures, while an accelerating expansion of the universe requires negative pressures.
Therefore, the understanding of cosmic acceleration turns out to be very complicated. This issue is commonly referred to as: the dark energy problem.
Recent measurements indicate that dark energy contributes 68\% of the total energy in the present-day observable universe.
In the last years, many different proposals have been suggested as theoretical frameworks for cosmic acceleration.
They can be divided basically in three main categories \cite{DE1}. A first one, in which the gravitational interaction
is modified (see e.g. \cite{DE1,odintsovreport,ODR1,ODR2,ODR3,ODR4,ODR5,ODR6,ODR7,ODR8,ODR9}), a second one in which the underlying geometry of the Universe is modified
(see e.g. \cite{MG1,MG2,MG3}) and a third one in which the models for the matter sector of the gravitational field equations
are refined and/or modified \cite{MMC1,MMC2,MMC3,MMC4,MMC5,MMC6,MMC7,MMC8,MMC9,MMC10}.
%
%
%
%
Other proposals rely on pure quantum field theoretical effects   and the non--trivial structure of the vacuum of flavor particle mixing \cite{CapDark1,CapDark2,CapDark3,CapDark4,CapDark5}.

The particle mixing and oscillations, that in the fermion sector characterize the evolution of neutrinos, and in the boson sector affect the dynamics of neutral kaons, $B^0$, $D^0$, and  $\eta-\eta'$ system, represent important phenomena of physics beyond the standard model of particles. Indeed, neutrino oscillations and $K^0$-$\overline{K}^0$ mixing play a crucial role in the analysis of the CP  violation and to test the CPT symmetry. Moreover, neutrinos, axions and axion-like particles which oscillate with photons \cite{Axion1,Axion2,Axion01,Axion02,Axion3,Axion03,Axion4,Axion04,Axion5,Axion6,Axion7,NeutDense} might constitute a component of the dark matter of the universe.
It has been also shown that, in flat space time,   the vacuum associated to the mixed particles, the flavor vacuum, behaves as a perfect fluid which has a state equation typical of the cold dark matter, in the case of fermion mixing, and a state equation of the dark energy, in the case of boson mixing \cite{CapDark1,CapDark2}.  The study of fermion mixing in curved space, together with the derivation of general oscillation formulae, has been carried out in refs.\cite{2020PhRvD.101i5022C,Buoninfante,Grossman,Piriz,Cardall}, and the analysis of the neutrino mixing contribution to the dark matter in curved background has been recently performed in ref.\cite{capolupoDarkMatter2022}.

Here, we consider mixed bosons in curved space. We extend the analysis presented in Refs.\cite{CapDark1,CapDark2,CapDark3,CapDark4,CapDark5}, in which the possible contribution of the flavor vacuum to the dark matter and energy was studied in Minkowski spacetime, and we analyze the behaviour of the bosonic flavor vacuum in the case of a curved background. In a previous work \cite{BosonOscillations} the fundamentals of the quantum field theory of boson mixing in curved space were laid out. Building on this formalism we take into account  a generic spatially flat Friedmann--Lemaitre--Robertson--Walker metric. We compute  the expectation value of the energy momentum tensor of free bosons on the vacuum for mixed fields and  we show that this tensor is  diagonal  and satisfies the Bianchi identities. These results are independent of the specific scale factor employed. Therefore, the bosonic flavor vacuum can be effectively considered as a perfect fluid.
In particular, assuming a fixed de Sitter background, we show that the  adiabatic factor $w^{(MIX)}$ due to the boson flavor mixing assumes values ranging from $-1$ to $1$. On the other hand, in the flat space time limit one has $w^{(MIX)}=-1$, which corresponds to the state equation of the dark energy.
The analysis here presented  gives an indication that the energy of the boson flavor vacuum may partially contribute to the cosmological dark energy. Elementary mixed bosons, leading to the aforementioned vacuum energy, may be represented by the neutrino super-partners.

The paper is organized as following. In section II, we describe the properties of the Klein-Gordon equation in curved space-time, focusing our attention on the Friedmann-Lemaitre-Robertson-Walker spacetime (FLRW). In section III,
 we  quantize a   boson field with definite mass, and   we introduce the  mixing of two flavor fields.
 The expectation value of the   stress energy tensor on the flavor vacuum  is derived in section IV, and, in section V we consider  the de Sitter space-time and  derive  the corresponding exact solution for the component of the stress energy tensor. Consideration on its regularization are also presented.
  Section VI is devoted to the conclusions.

\section{Klein-Gordon fields in flat FLRW spacetime}

We  study the cosmological effects of particle mixing in curved space-time. In particular, we  focus our attention on the spatially flat Friedmann-Lema\^{\i}tre-Robertson-Walker spacetime (FLRW), $ds^2 = dt^2 - C^2(t) d \textbf{x}^2$, where $C (t)$ is the scale factor, since the FLRW metric well describes the expansion of the universe in the present epoch. We start this section by   summarizing the solution of Klein-Gordon equation in the FLRW spacetimes and use the $(+,-,-,-)$ signature, so that the metric tensor is
$
 g_{\mu \nu} = \mathrm{diag}\left(1,-C^2(t),-C^2(t),-C^2(t)\right).
$
  For our purposes, it is useful to express this metric in terms of the conformal time $\tau$, defined as
$
d\tau = \frac{dt}{C(t)}\,.
$
The range of the conformal time $\tau $ corresponding to the coordinate time interval $t \in \mathbb{R} $ depends on the specific scale factor considered $C(t)$.
In terms of $\tau$, the line element reads
\begin{equation}\label{LineElement}
  ds^2 = C^2(\tau) [d\tau^2 - dx^2 -dy^2 -dz^2]\,.
\end{equation}

We consider two free charged scalar fields $\phi_i$ with masses $m_i$, for $i=1,2$, with the minimally coupled Lagrangian ($\xi = 0$)
\begin{equation}\label{Lagrangian}
  \mathcal{L} = \sum_{i=1,2} \left \lbrace \frac{1}{2}\sqrt{-g} \left[g^{\mu \nu} \partial_{\mu}\phi^{\dagger}_i \partial_{\nu}\phi_i -m^{2}_i\phi^{\dagger}_i\phi_i \right] \right \rbrace \,,
\end{equation}
from which the Klein-Gordon equations
$
  (\square+m^{2}_i)\phi_i = 0 \, ,
$
are derived. The symbol $\square$ stands here for the curved space D'alembertian.
The scalar product between two solutions $A$ and $ B$ of the Klein-Gordon equation is defined as
\begin{equation}\label{ScalarProduct}
  (A,B)_{\tau}=i\int_{\Sigma_{\tau}}d \ \Sigma^{\mu} \sqrt{-g}\left(A^{*}\partial_{\mu} B-\left(\partial_{\mu}A^{*}\right) B\right)=i\int_{\Sigma_{\tau}}d^3x \sqrt{-g} g^{\tau \tau}\left(A^{*}\partial_{\tau} B-\left(\partial_{\tau}A^{*}\right) B\right) \ ,
\end{equation}
where the last equality holds for hypersurfaces $\Sigma_{\tau}$ of constant conformal time $\tau$, on which the integration is to be performed. If $A$ and $B$ are solutions of the \textit{same} Klein-Gordon equation, the scalar product $(A,B)_{\tau}$ is independent of $\tau$. In general this does not hold true for solutions to Klein-Gordon equations with distinct masses.

The energy-momentum tensor is obtained as usual by varying the action $S= \int d^4 x \mathcal{L}$ with respect to the metric; it is given by
\begin{equation}\label{EnergyPulseTensor1}
T_{\mu \nu}=\frac{1}{2} \sum_{i=1,2} \left \lbrace \partial_{\mu} \phi^{\dagger}_i \partial_{\nu} \phi_i +\partial_{\nu} \phi^{\dagger}_i \partial_{\mu} \phi_i -g_{\mu \nu} g^{\rho \sigma} \partial_{\rho} \phi^{\dagger}_i \partial_{\sigma} \phi_i + m^2_i g_{\mu \nu} \phi^{\dagger}_i\phi_i\right \rbrace
\end{equation}

\section{Quantization of flavor fields}

In the section, we first  quantize a single boson field of definite mass, and then we  analyze   the main properties of two flavor mixed fields.

\subsection{Boson field }

Let us consider two complete sets of solutions to the Klein-Gordon equations with masses $m_i$ $\lbrace{ u_{\pmb{p};i},u^{*}_{\pmb{p};i} \rbrace}$. The notation here anticipates that, given the general form of the metric, the spatial part of the equations is solved by plane waves labelled by $3$-vectors $\pmb{p}$. Any solution of the (linear) Klein-Gordon equations can be written as a linear combination of these modes. In particular, the boson fields can be expressed as:
$\label{FieldExpansion}
  \phi_i (x) = \int d^3 p \left(a_{\pmb{p};i} u_{\pmb{p};i} + b^{*}_{-\pmb{p};i} u^{*}_{-\pmb{p};i} \right)
$,
where  the spacetime dependence is contained in the modes, while the coefficients are independent of the space and time coordinates.
Quantization proceeds in analogy to the Minkowski space, by promoting the fields $\phi_i$ to operators
\begin{equation}\label{FieldExpansion2}
  \phi_i (x) =\int d^3 p \left(a_{\pmb{p};i} u_{\pmb{p};i} + b^{\dagger}_{-\pmb{p};i} u^{*}_{-\pmb{p};i} \right) \ ,
\end{equation}
and imposing the canonical commutation relations between  $\phi_i(x)$ and its conjugate momentum. The choice of the opposite label ($-\pmb{p}$) for the antiparticles is just a matter of convention, but it gives to both the terms of the expansion in Eq. \eqref{FieldExpansion2} the same spatial dependency.

This translates to the following commutation relations for the creation and annihilation operators:
\begin{equation}
  \left[ a_{\pmb{p};i}, a^{\dagger}_{\pmb{q};j} \right] =  \left[ b_{\pmb{p};i}, b^{\dagger}_{\pmb{q};j} \right] = \delta_{ij} \delta^3 (\pmb{p}-\pmb{q}),
\end{equation}
with all the other commutators vanishing.  The vacuum state $\ket{0}$ is then defined as
$
  a_{\pmb{p};i} \ket{0} = b_{\pmb{p};i} \ket{0}= 0 \,$ for all $\pmb{p}\
$ and $i$.
By introducing the field expansion in Eq.(\ref{EnergyPulseTensor1}), one obtains the quantized energy-momentum tensor of the boson fields:
\begin{eqnarray}\label{EnergyPulseTensor2}
 \nonumber  T_{\mu \nu}&=& \sum_{i=1,2} \int d^3 p \int d^3 q \lbrace a^{\dagger}_{\pmb{p};i}a_{\pmb{q};i} L_{\mu \nu} (u_{\pmb{p};i},u_{\pmb{q};i}) + a^{\dagger}_{\pmb{p};i}b^{\dagger}_{-\pmb{q};i} L_{\mu \nu} (u_{\pmb{p};i},u^{*}_{-\pmb{q};i}) \\
&+& b_{-\pmb{p};i}a_{\pmb{q};i} L_{\mu \nu} (u^{*}_{-\pmb{p};i},u_{\pmb{q};i}) + b_{-\pmb{p};i}b^{\dagger}_{-\pmb{q};i} L_{\mu \nu} (u^{*}_{-\pmb{p};i},u^{*}_{-\pmb{q};i})\rbrace \ .
\end{eqnarray}

In Eq.\eqref{EnergyPulseTensor2} there is a neat separation between the operator part and the functional part, represented by the tensor functional $L_{\mu \nu} (A,B)$, defined on the solutions $A,B$ of the Klein Gordon equations. By definition one has
\begin{equation}\label{AuxiliaryTensorDefinition}
 L_{\mu \nu} (A_i , B_i) = \frac{1}{2}  \left \lbrace \partial_{\mu} A^{*}_i \partial_{\nu} B_i +\partial_{\nu} A^{*}_i \partial_{\mu} B_i -g_{\mu \nu} g^{\rho \sigma} \partial_{\rho} A^{*}_i \partial_{\sigma} B_i + m^2_i g_{\mu \nu} A^{*}_i B_i\right \rbrace
\end{equation}
for any two solutions $A_i, B_i$ of the Klein-Gordon equation with mass $m_i$. Some elementary properties of $L_{\mu \nu}$ can be read off the definition. Clearly $L_{\mu \nu}$ is symmetric ($=L_{\nu \mu}$) for any argument and
\begin{equation}\label{Lproperty1}
 L_{\mu \nu} (A_i, B_i) = L^*_{\mu \nu} (B_i, A_i) \ .
\end{equation}
In particular Eq. \eqref{Lproperty1} implies that $L_{\mu \nu} (A,A)$ is real for any $A$. Additional properties of $L_{\mu \nu}$ are shown below.

\subsection{The flavor fields}

We now move on to the quantization of the flavor fields, essentially following the treatment in \cite{BosonOscillations}. As usual, the flavor fields are  defined by means of the rotation
\begin{eqnarray}\nonumber
  \phi_{A}(x) &=& \cos(\theta) \phi_{1}(x) + \sin(\theta) \phi_2(x)
  \\
  \ \phi_{B}(x) &=& \cos(\theta) \phi_{2}(x) - \sin(\theta) \phi_1(x) \ ,
\end{eqnarray}
where $\theta$ is the  mixing angle. The rotation can be recast in terms of the mixing   generator
\begin{equation}\label{Generator}
  \mathcal{G}_{\theta} (\tau)= exp\left \lbrace\theta \left[ (\phi_1,\phi_2)_{\tau} - (\phi_2,\phi_1)_{\tau} \right] \right \rbrace \ ,
\end{equation}
where $(\phi_2,\phi_1)_{\tau}$ is  the scalar product at the $\tau$ hypersurface as defined in Eq.\eqref{ScalarProduct}. Then, the flavor fields can be expressed as
\begin{eqnarray}\nonumber
  \phi_{A}(x) &=& \mathcal{G}^{-1}_{\theta}\,  \phi_1 (x)\, \mathcal{G}_{\theta}
   \\
   \phi_{B}(x) &=& \mathcal{G}^{-1}_{\theta}\,  \phi_2 (x)\, \mathcal{G}_{\theta} \ .
\end{eqnarray}
In a similar way, the flavor annihilators are defined as $a_{\pmb{p};\sigma}  = \mathcal{G}^{-1}_{\theta}\,  a_{\pmb{p};j}\, \mathcal{G}_{\theta} $, with $\sigma = A, B$ and $j=1,2$, and similar for the antiparticles. The flavor vacuum, annihilated by   the flavor annihilators, is given by
\begin{equation}
  \ket{0_F (\tau)} = \mathcal{G}^{-1}_{\theta} (\tau) \ket{0} \ ,
\end{equation}
where $\ket{0}$ is the vacuum defined by the mass annihilators. Notice that, $\ket{0_F (\tau)}$ carries an explicit $\tau$ dependence due to $\mathcal{G}^{-1}_{\theta}(\tau)$.

\section{VEV of the energy-momentum tensor on the flavor vacuum}

Here, we compute the contribution of the flavor vacuum to the energy and to the pressure. To do that, we calculate the expectation value of the energy-momentum tensor at time  $\tau$ on the flavor vacuum at a given fixed time $\ket{0_F(\tau_0)}$, with $\tau_0 $ not necessarily coincident with the time argument of the energy momentum tensor $\tau$. We  consider a specific expansion of the mass fields, and thus a specific choice of the mass vacuum as suggested by the form of the metric. Other mass representations are of course possible, and the effect of a change of mass representation on the flavor fields can be obtained by the adequate transformations described in \cite{BosonOscillations}.
 The quantity we wish to compute is

\begin{equation}\label{VeV}
 \mathbb{T}_{\mu \nu} = \bra{0_F(\tau_0)} T_{\mu \nu} \ket{0_F(\tau_0)} \ ,
\end{equation}
where $T_{\mu \nu}$ is given by Eq.(\ref{EnergyPulseTensor2}).
We start by considering the typical term in Eq.(\ref{VeV}), which has the form
\begin{equation}
  \bra{0_F(\tau_0)} a^{\dagger}_{\pmb{p};1} a_{\pmb{q};1} \ket{0_F(\tau_0)} \ .
\end{equation}
By using the definition of the flavor vacuum such expectation value can be written as
\begin{eqnarray}
 \nonumber \bra{0}\mathcal{G}_{\theta}(\tau_0) a^{\dagger}_{\pmb{p};1} a_{\pmb{q};1}\mathcal{G}^{-1}_{\theta}(\tau_0) \ket{0} &= &  \nonumber  \bra{0}\mathcal{G}_{\theta}(\tau_0) a^{\dagger}_{\pmb{p};1} \mathcal{G}^{-1}_{\theta}(\tau_0) \mathcal{G}_{\theta}(\tau_0) a_{\pmb{q};1}\mathcal{G}^{-1}_{\theta}(\tau_0) \ket{0}
 \\ &=&
\bra{0}\mathcal{G}^{-1}_{-\theta}(\tau_0) a^{\dagger}_{\pmb{p};1} \mathcal{G}_{-\theta}(\tau_0) \mathcal{G}^{-1}_{-\theta}(\tau_0) a_{\pmb{q};1}\mathcal{G}_{-\theta}(\tau_0) \ket{0}
\end{eqnarray}
where we have used the relation (Eq. \ref{Generator}) $\mathcal{G}^{-1}_{\theta} = \mathcal{G}_{-\theta} $.
The transformed operator  $\mathcal{G}^{-1}_{-\theta}(\tau_0) a_{\pmb{p};1} \mathcal{G}_{-\theta}(\tau_0)$ and the others relative to the mass $2$ are just the mass annihilators transformed according to the mixing transformation with angle $-\theta$. Such annihilators are given by:

\begin{equation}\label{MixingTransformations}
\begin{split}
\mathcal{G}_{-\theta}^{-1} (\tau_0) a_{\pmb{p};1} \mathcal{G}_{-\theta}(\tau_0) &= \cos(\theta)a_{\pmb{p};1}-\sin(\theta) \left(\Lambda^*_{\pmb{p}}(\tau_0) a_{\pmb{p};2} + \Xi_{\pmb{p}}(\tau_0)b^{\dagger}_{-\pmb{p};2} \right) \\
\mathcal{G}_{-\theta}^{-1} (\tau_0) a_{\pmb{p};2} \mathcal{G}_{-\theta}(\tau_0) &= \cos(\theta)a_{\pmb{p};2}+\sin(\theta) \left(\Lambda_{\pmb{p}}(\tau_0) a_{\pmb{p};1} - \Xi_{\pmb{p}}(\tau_0)b^{\dagger}_{-\pmb{p};1} \right) \\
 \mathcal{G}_{-\theta}^{-1} (\tau_0) b_{-\pmb{p};1} \mathcal{G}_{-\theta}(\tau_0) &= \cos(\theta)b_{-\pmb{p};1}-\sin(\theta) \left(\Lambda^*_{\pmb{p}}(\tau_0) b_{-\pmb{p};2} + \Xi_{\pmb{p}}(\tau_0)a^{\dagger}_{\pmb{p};2} \right) \\
 \mathcal{G}_{-\theta}^{-1} (\tau_0) b_{-\pmb{p};2} \mathcal{G}_{-\theta}(\tau_0) &= \cos(\theta)b_{-\pmb{p};2}+\sin(\theta) \left(\Lambda_{\pmb{p}}(\tau_0) b_{-\pmb{p};1} - \Xi_{\pmb{p}}(\tau_0)a^{\dagger}_{\pmb{p};1} \right)
\end{split}
\end{equation}
 The Bogoliubov coefficients are defined by means of the inner products
\begin{equation}\label{BogoliubovCoefficients}
\begin{split}
 \delta^3 (\pmb{p} - \pmb{q})\Lambda_{\pmb{p}}(\tau) &= \left(u_{\pmb{p};2},u_{\pmb{q};1}\right)_{\tau} \\
  \delta^3 (\pmb{p} + \pmb{q})\Xi_{\pmb{p}}(\tau) &= \left(u_{\pmb{p};1},u^{*}_{\pmb{q};2}\right)_{\tau}\ ,
  \end{split}
\end{equation}
where the delta function is absorbed by a corresponding momentum integration in the Eqs.(\ref{MixingTransformations}). For distinct labels $\pmb{p}, \pmb{q}$, the inner products vanish. The coefficients in Eq.(\ref{BogoliubovCoefficients})  satisfy the condition $|\Lambda_{\pmb{p}} |^2 - |\Xi_{\pmb{p}} |^2=1$ for all $\pmb{p},\tau$. By using Eqs.(\ref{MixingTransformations}), we can compute  the expectation values present in Eq.(\ref{VeV}):
\begin{equation}
\begin{split}
  \bra{0_F(\tau_0)}a^{\dagger}_{\pmb{p};j}a_{\pmb{q};j}\ket{0_F(\tau_0)} &= \sin^2 \theta \, |\Xi_{\pmb{p}}(\tau_0)|^2  \, \delta^3 (\pmb{p}-\pmb{q})\ \,, \ \ \ \forall j \\
   \bra{0_F(\tau_0)}b^{\dagger}_{-\pmb{p};j}b_{-\pmb{q};j}\ket{0_F(\tau_0)} &= \sin^2 \theta  \,  |\Xi_{\pmb{p}}(\tau_0)|^2  \,  \delta^3 (\pmb{p}-\pmb{q})\ \,, \ \ \ \forall j \\
 \bra{0_F(\tau_0)}a^{\dagger}_{\pmb{p};1}b^{\dagger}_{-\pmb{q};1}\ket{0_F(\tau_0)} &=\sin^2 \theta  \,  \Xi^*_{\pmb{p}}(\tau_0) \Lambda_{\pmb{p}}(\tau_0)  \,  \delta^3 (\pmb{p}-\pmb{q}) \\
 \bra{0_F(\tau_0)}a^{\dagger}_{\pmb{p};2}b^{\dagger}_{-\pmb{q};2}\ket{0_F(\tau_0)} &= -\sin^2 \theta  \,  \Xi^*_{\pmb{p}}(\tau_0) \,  \Lambda^*_{\pmb{p}} \, (\tau_0) \delta^3 (\pmb{p}-\pmb{q}) \\
 \bra{0_F(\tau_0)}b_{-\pmb{p};1}a_{\pmb{q};1}\ket{0_F(\tau_0)} &= \sin^2 \theta \,  \Xi_{\pmb{p}}(\tau_0)  \, \Lambda^*_{\pmb{p}}(\tau_0) \, \delta^3 (\pmb{p}-\pmb{q}) \\
 \bra{0_F(\tau_0)}b_{-\pmb{p};2}a_{\pmb{q};2}\ket{0_F(\tau_0)} &= -\sin^2 \theta  \, \Xi_{\pmb{p}}(\tau_0) \Lambda_{\pmb{p}}(\tau_0)  \, \delta^3 (\pmb{p}-\pmb{q}) \ . \\
\end{split}
\end{equation}
The expectation value of the energy momentum tensor is then given by two contributions as follows:
\begin{eqnarray}\label{VeV2}
\mathbb{T}_{\mu \nu} & =& \mathbb{T}_{\mu \nu}^{(MIX)} + \mathbb{T}_{\mu \nu}^{(N)} \\
\nonumber\mathbb{T}_{\mu \nu}^{(MIX)} &=&  \sin^2 \theta \,\int d^3 p \bigg\lbrace |\Xi_{\pmb{p}}(\tau_0)|^2 \sum_{j=1,2} \left( L_{\mu \nu}(u_{\pmb{p};j},u_{\pmb{p};j}) + L_{\mu \nu}(u^*_{-\pmb{p};j},u^{*}_{-\pmb{p};j})\right) \\
\nonumber & +& \ \Xi^*_{\pmb{p}}(\tau_0)\Lambda_{\pmb{p}}(\tau_0) L_{\mu \nu}(u_{\pmb{p};1},u^{*}_{-\pmb{p};1}) + \Xi_{\pmb{p}}(\tau_0) \Lambda^*_{\pmb{p}}(\tau_0) L_{\mu \nu}(u^*_{-\pmb{p};1},u_{\pmb{p};1}) \\
 & - &\Xi^*_{\pmb{p}}(\tau_0)\Lambda^*_{\pmb{p}}(\tau_0) L_{\mu \nu}(u_{\pmb{p};2},u^{*}_{-\pmb{p};2}) - \Xi_{\pmb{p}}(\tau_0) \Lambda_{\pmb{p}}(\tau_0) L_{\mu \nu}(u^*_{-\pmb{p};2},u_{\pmb{p};2}) \bigg\rbrace \\
\mathbb{T}_{\mu \nu}^{(N)} & =&  \sum_{j=1,2}\int d^3 p L_{\mu \nu}(u^*_{-\pmb{p};j},u^{*}_{-\pmb{p};j}) \ .
\end{eqnarray}
Here the first term is exclusively due to the mixing, indeed $\mathbb{T}_{\mu \nu}^{(MIX)}$   depends on $\sin^2 \theta$ and vanishes for $\theta=0$. The last term derives from  the commutation relation $\left[ b_{-\pmb{p};j}, b^{\dagger}_{-\pmb{q;j}} \right] = \delta^3(\pmb{p}-\pmb{q})$ applied to the $bb^{\dagger}$ term.
    $\mathbb{T}^{(N)}_{\mu \nu}$ is the expectation value of the energy-momentum tensor on the \emph{mass} vacuum:
\begin{equation}
  \mathbb{T}^{(N)}_{\mu \nu} = \bra{0} T_{\mu \nu} \ket{0} \ .
\end{equation}

The $(0,0)$ component of this term corresponds to the diverging energy that is removed by the normal ordering  in flat space.
Indeed, in the flat space, one has $u^{*}_{-\pmb{p}; j}=\frac{1}{\sqrt{2(2\pi)^{3}\omega_{p;j}}} e^{i \omega_{p;j} t+i\pmb{p}\cdot \pmb{x}}$ with $\omega_{p;j} = \sqrt{p^2 + m_j^2}$, and  the auxiliary tensor becomes
\begin{equation}
 L_{\mu \nu} (u^*_{-\pmb{p};j},u^{*}_{-\pmb{p};j}) = \partial_{\mu}u_{-\pmb{p};j} \partial_{\nu} u^{*}_{-\pmb{p};j} + \partial_{\nu}u_{-\pmb{p};j}\partial_{\mu} u^{*}_{-\pmb{p};j} - \eta_{\mu\nu}\partial_{\rho} u_{-\pmb{p};j}\partial^{\rho} u^*_{-\pmb{p};j} +m^{2}\eta_{\mu\nu}|u_{-\pmb{p};j}|^2
\end{equation}
where  the metric tensor and the derivatives are those of the ordinary flat space. For the $(0,0)$ component,   we have
\begin{equation}
 L_{00} (u^*_{-\pmb{p};j},u^{*}_{-\pmb{p};j}) = \frac{\omega_{p;j}}{2}  \ ,
\end{equation}
 then
\begin{equation}
  \mathbb{T}^{(N)}_{0 0} = \sum_{j=1,2} \int d^3p \  \frac{\omega_{p;j}}{2} \ \ \ \ \ \ \ (>0)     \  .
\end{equation}
 Clearly this term has to be removed if the curved space energy momentum tensor is to approach the normal ordered energy momentum tensor of flat space in the appropriate limit. This property is featured also among the Wald's axioms for the energy-momentum tensors in curved space \cite{Wald}. We then define the renormalized energy-momentum tensor as
\begin{equation}
 T_{\mu \nu}^r = T_{\mu \nu} - \mathbb{T}^{(N)}_{\mu \nu}\,.
\end{equation}
Its expectation value is then
\begin{equation}
 \bra{0_F(\tau_0)}T_{\mu \nu}^r\ket{0_F(\tau_0)} = \mathbb{T}^{(MIX)}_{\mu \nu}\ .
\end{equation}
In the following we show the main properties of the Bogoliubov coefficients and we demonstrate that $\mathbb{T}_{\mu \nu}$ corresponds to the energy momentum tensor of a perfect fluid.

\subsection{General properties of the Bogoliubov Coefficients}

Given the form of the Klein-Gordon equations, we employ the following ansatz for the solutions:
\begin{equation}\label{SolutionAnsatz}
 u_{\pmb{p};j} (\tau, \pmb{x}) = (2\pi)^{-\frac{3}{2}}e^{i \pmb{p} \cdot \pmb{x}} C^{-1}(\tau) \chi_{p,j} (\tau) \ .
\end{equation}
The functions $\chi_{p,j} (\tau)$ depend only on the modulus of the momentum $p = |\pmb{p}|$ and on the conformal time $\tau$. Inserting the ansatz \eqref{SolutionAnsatz} in the inner products defining the Bogoliubov coefficients of Eq. \eqref{BogoliubovCoefficients}, and recalling the form of the metric in conformal time, we obtain

\begin{eqnarray}\label{BogoliubovCoefficients2}
\nonumber \Lambda_p (\tau) &=& i \left( \chi_{p;2}^*(\tau) \partial_{\tau} \chi_{p,1} (\tau) -\left( \partial_{\tau} \chi^*_{p,2} (\tau) \right)  \chi_{p;1}(\tau) \right) \ \\
\Xi_p (\tau) &=& i \left( \chi_{p;1}^*(\tau) \partial_{\tau} \chi_{p,2}^* (\tau) -\left( \partial_{\tau} \chi^*_{p,1} (\tau) \right)  \chi_{p;2}(\tau) \right) \ .
\end{eqnarray}

It can be easily checked that the fundamental property
 \begin{eqnarray}\label{BogoliubovCoefficients4}
|\Lambda_{p}(\tau)|^2 - |\Xi_p (\tau)|^2 &=&
1 \ ,
\end{eqnarray}
holds, provided that the normalization $(u_{\pmb{p};i},u_{\pmb{q};j}) = \delta_{ij}\delta^{(3)}(\pmb{p}-\pmb{q}) = -(u^*_{\pmb{p};i},u^*_{\pmb{q};j})$ is used. For completeness, on the reduced modes the normalization condition reads

\begin{equation}
 i \left( \chi_{p;j}^*(\tau) \partial_{\tau} \chi_{p,j} (\tau) -\left( \partial_{\tau} \chi^*_{p,j} (\tau) \right)  \chi_{p;j}(\tau) \right) = 1 \ ; \ \  \ \ \ \ \ \ \ \ \forall j=1,2.
\end{equation}

\subsection{Diagonality of the energy-momentum tensor}

We now show that $\mathbb{T}_{\mu\nu}$ can be interpreted as the energy-momentum tensor of a perfect fluid. This result relies on the properties of the auxiliary tensor $L_{\mu \nu}$ in a spatially flat and isotropic metric. We first show that
\begin{equation}
 L_{\tau i} (u_{\pmb{p};j}, u_{\pmb{p};j}) = p_i f_1(p,\tau) ; \ \ \ \  L_{\tau i} (u_{\pmb{p};j}, u^{*}_{-\pmb{p};j}) = p_i f_2(p,\tau); \ \ \ \  \forall j=1,2; i = 1,2,3
\end{equation}
where $f_{1,2}$ are functions of the modulus $p$ and of the conformal time $\tau$ alone. Let us insert the ansatz \eqref{SolutionAnsatz} in the definition \eqref{AuxiliaryTensorDefinition}:
\begin{eqnarray}
 \nonumber L_{\tau i} (u_{\pmb{p};j},u_{\pmb{p};j}) &=& \frac{1}{2} \left \lbrace \partial_{\tau} u^*_{\pmb{p};j} \partial_i u_{\pmb{p};j} + \partial_{i} u^{*}_{\pmb{p};j} \partial_{\tau} u_{\pmb{p};j} \right \rbrace \\ \nonumber &=& \frac{1}{2 (2 \pi)^3} \left \lbrace i p_i \left(\dot{C}C^{-3} \chi^*_{p,j}(\tau) + C^{-2} \dot{\chi}^*_{p;j} (\tau) \right)\chi_{p;j} (\tau) - ip_i \chi^*_{p;j}(\tau)\left(\dot{C}C^{-3} \chi_{p,j}(\tau) + C^{-2} \dot{\chi}_{p;j} (\tau) \right)\right \rbrace \\
 \nonumber &=& p_i \left \lbrace \frac{i}{2 (2\pi)^3} \left[\left(\dot{C}C^{-3} \chi^*_{p,j}(\tau) + C^{-2} \dot{\chi}^*_{p;j} (\tau) \right) \chi_{p;j}(\tau) - \chi^*_{p;j}(\tau)\left(\dot{C}C^{-3} \chi_{p,j}(\tau) + C^{-2} \dot{\chi}_{p;j} (\tau) \right) \right] \right \rbrace \ .
\end{eqnarray}
The parenthetical term clearly depends only on $p$ and $\tau$. Here the dot denotes a derivative with respect to $\tau$. Similarly
\begin{equation*}
 \nonumber L_{\tau i} (u_{\pmb{p};j},u*_{-\pmb{p};j}) = p_i \left \lbrace \frac{-i}{ (2\pi)^3} \left[\left(\dot{C}C^{-3} \chi^*_{p,j}(\tau) + C^{-2} \dot{\chi}^*_{p;j} (\tau) \right) \chi^*_{p;j}(\tau)  \right] \right \rbrace \ .
\end{equation*}
Due to the basic property $\eqref{Lproperty1}$, the auxiliary tensor has the same form also for the other arguments. Because the Bogoliubov coefficients depend only on $p$ and $\tau$, the overall form of $\mathbb{T}^{(MIX)}_{\tau i}$ is
\begin{equation}
 \mathbb{T}^{(MIX)}_{\tau i} = \int d^3 p \  p_i F(p,\tau)
\end{equation}
for some function $F$ of $p$ and $\tau$. This quantity clearly vanishes due to symmetry (the $p_i$ integral extends over the whole of $\mathbb{R}$). A similar argument applies to the spatial components, for which

\begin{equation}
 L_{ik} (u_{\pmb{p};j}, u_{\pmb{p};j}) = p_i p_k g_1(p,\tau) ; \ \ \ \  L_{ik} (u_{\pmb{p};j}, u^{*}_{-\pmb{p};j}) = p_i p_k g_2(p,\tau); \ \ \ \  \forall j=1,2; i,k = 1,2,3 \ i \neq k .
\end{equation}
Indeed
\begin{eqnarray}
\nonumber L_{ik}(u_{\pmb{p};j}, u_{\pmb{p};j}) &=& \frac{1}{2} \left \lbrace \partial_{i} u^*_{\pmb{p};j} \partial_{k} u_{\pmb{p};j}  + \partial_{k} u^{*}_{\pmb{p};j} \partial_{i} u_{\pmb{p};j}\right \rbrace \\
\nonumber &=& \frac{1}{2 (2\pi)^3} \left \lbrace 2 p_i p_k C^{-2} |\chi_{p;j}(\tau)|^2 \right \rbrace = p_i p_k \left \lbrace \frac{|\chi_{p;j}(\tau)|^2}{(2\pi)^3 C^2}  \right \rbrace
\end{eqnarray}
and
\begin{equation*}
  L_{ik}(u_{\pmb{p};j}, u_{-\pmb{p};j}) = p_i p_k \left \lbrace \frac{(\chi^*_{p;j}(\tau))^2}{(2 \pi)^3 C^2} \right \rbrace
\end{equation*}
so that overall
\begin{equation}
 \mathbb{T}^{(MIX)}_{ik} = \int d^3 p \ p_i  p_k G(p, \tau)
\end{equation}
for some function $G(p,\tau)$ of $p$ and $\tau$ alone. For $i \neq k$ the integral is zero due to symmetry. This proves that $\mathbb{T}_{\mu \nu}$ is diagonal. It is likewise easy to show that for $i = k$ one has
\begin{equation}
 \mathbb{T}^{(MIX)}_{ii} = \int d^3 p \  H(p,\tau)
\end{equation}
for some function $H$ of $p$ and $\tau$. As an immediate consequence, and as expected from the isotropy of the metric, the tensor is also isotropic, i. e. $\mathbb{T}^{(MIX)}_{ii}$ is the same for all $i=1,2,3$. Below we shall compute explicitly the two non-zero components of $\mathbb{T}_{\mu \nu}$. We shall see that the only residual dependency on the coordinates is on the conformal time $\tau$, while no spatial dependency arises. Then the vacuum expectation value respects the spatial translation symmetry of the metric.

\subsection{Energy density and pressure}

Another trivial property of the auxiliary tensor, that can be read straight off the definition \eqref{AuxiliaryTensorDefinition}, is
\begin{equation}\label{Lproperty2}
 L_{\mu \nu} (A_i^* , B_i^*) = L_{\mu \nu}^* (A_i, B_i) \,\, \Longrightarrow \,\,  L_{\mu \nu} (A_i^*, A_i^*) = L^*_{\mu \nu} (A_i, A_i) = L_{\mu \nu} (A_i, A_i) \ .
\end{equation}
In the last equality we have taken into account the reality of $L_{\mu \nu} (A_i, A_i)$ (see Eq. \eqref{Lproperty1}). Together with Eq. \eqref{Lproperty1} this allows us to write
\begin{eqnarray}\label{ReducedEnergyMomentumTensor}
 \nonumber \mathbb{T}_{\mu \nu}^{(MIX)} &=& \sin^2 \theta \int d^3 p \bigg \lbrace 2 |\Xi_p (\tau_0)|^2 \sum_{j=1,2} L_{\mu \nu} (u_{\pmb{p};j}, u_{\pmb{p};j}) + \left[ \Xi_p^* (\tau_0) \Lambda_p (\tau_0) L_{\mu \nu} (u_{\pmb{p};1},u^*_{-\pmb{p};1}) + c.c. \right] \\
 &-& \left[\Xi_p (\tau_0), \Lambda_p (\tau_0) L^*_{\mu \nu} (u_{\pmb{p};2}, u^*_{-\pmb{p};2}) + c.c. \right] \bigg \rbrace \ .
\end{eqnarray}
As shown above only the diagonal components are non-zero, with the three spatial components identified. We then need only the following components of the auxiliary tensor:

\begin{equation}\label{AuxTensC1}
 L_{\tau \tau} (u_{\pmb{p};j}, u_{\pmb{p};j}) = \frac{1}{2 (2\pi)^3} \left \lbrace |\chi_{p;j} |^2 \left(p^2 C^{-2} + m_j^2 + C^{-4} \dot{C}^2  \right) + C^{-2}  |\dot{\chi}_{p;j}|^2 - C^{-3} \dot{C} \left(\chi^*_{p;j} \dot{\chi}_{p;j} + \dot{\chi}^*_{p;j}\chi_{p:j} \right) \right \rbrace
\end{equation}

\begin{equation}\label{AuxTensC2}
 L_{\tau \tau} (u_{\pmb{p};j}, u^*_{-\pmb{p};j}) = \frac{1}{2(2\pi)^3} \left \lbrace (\chi^*_{p;j} )^2 \left(p^2 C^{-2} + m_j^2 + C^{-4} \dot{C}^2  \right) + C^{-2}  (\dot{\chi}^*_{p;j})^2 - 2C^{-3} \dot{C} \chi^*_{p;j} \dot{\chi}^*_{p;j}  \right \rbrace
\end{equation}

\begin{equation}\label{AuxTensC3}
 L_{kk} (u_{\pmb{p};j}, u_{\pmb{p};j}) = \frac{1}{2(2\pi)^3} \left \lbrace |\chi_{p;j} |^2 \left(2p_k^2 C^{-2} + C^{-4}\dot{C}^2 - p^2 C^{-2} -m_j^2\right) + C^{-2}|\dot{\chi}_{p;j}|^2 -C^{-3}\dot{C}\left( \chi^*_{p;j}\dot{\chi}_{p;j} +\dot{\chi}^*_{p;j} \chi_{p;j}\right)\right \rbrace
\end{equation}

\begin{equation}\label{AuxTensC4}
 L_{kk} (u_{\pmb{p};j}, u^*_{-\pmb{p};j}) = \frac{1}{2(2\pi)^3} \left \lbrace (\chi^*_{p;j} )^2 \left(2p_k^2 C^{-2} + C^{-4}\dot{C}^2 - p^2 C^{-2} -m_j^2\right) + C^{-2}(\dot{\chi}^*_{p;j})^2 -2C^{-3}\dot{C} \chi^*_{p;j}\dot{\chi}^*_{p;j} \right \rbrace \ .
\end{equation}

Here $k =1,2,3$ denotes any of the spatial indices. In all of the above equations it is understood that the mode functions $\chi_{p;j} \equiv \chi_{p;j}(\tau)$ and the scale factor $C \equiv C(\tau)$ depend on the conformal time $\tau$ alone. Given that the Bogoliubov coefficients in Eq. \eqref{ReducedEnergyMomentumTensor} depend only on the reference time $\tau_0$, it is clear, as claimed above, that the diagonal components of $\mathbb{T}_{\mu \nu}^{(MIX)}$ depend only on the conformal time $\tau$ and the reference time $\tau_0$. Due to the form of $\mathbb{T}_{\mu \nu}^{(MIX)}$ we can identify the energy density associated to the flavor vacuum $ \rho^{(MIX)} (\tau_0,\tau) $ with the $\tau \tau$ component of $\mathbb{T}_{\mu}^{(MIX) \nu}$ and the corresponding pressure $p^{(MIX)} (\tau_0,\tau)$ with $\mathbb{T}_{k}^{(MIX) k}$:
\begin{equation}
 \rho^{(MIX)} (\tau_0,\tau) = \mathbb{T}_{\tau}^{(MIX) \tau} = g^{\tau \tau} \mathbb{T}^{(MIX)}_{\tau \tau } = C^{-2} (\tau) \mathbb{T}^{(MIX)}_{\tau \tau} (\tau_0, \tau)
\end{equation}
\begin{equation}
 p^{(MIX)} (\tau_0,\tau) = -\mathbb{T}_{k}^{(MIX) k} = -g^{k k} \mathbb{T}^{(MIX)}_{k k } = C^{-2} (\tau) \mathbb{T}^{(MIX)}_{k k} (\tau_0, \tau).
\end{equation}
Here no sum over $k$ is intended.

\subsection{Minkowskian Limit}

Let us compute the energy density and pressure in the flat space limit $C(t) \rightarrow 1$, $\tau \rightarrow t$. The reduced modes of Eq. \eqref{SolutionAnsatz} take the usual form
\begin{equation}
 \chi_{p;j} (t) = \frac{1}{\sqrt{2 \omega_{p;j}}} e^{-i \omega_{p;j}t}, \ \ \ \ \omega_{p;j} = \sqrt{p^2 + m_j^2} \ .
\end{equation}
Insertion in Eqs. \eqref{BogoliubovCoefficients2} yields the known flat spacetime expressions
\begin{equation}
 \Lambda_p (t_0) = \frac{(\omega_{p;1} + \omega_{p;2})}{\sqrt{4 \omega_{p;1} \omega_{p;2}}}  e^{i (\omega_{p;2}-\omega_{p;1})t_0} \ \ ; \ \ \ \ \ \Xi_p (t_0) = \frac{(\omega_{p;1} - \omega_{p;2})}{\sqrt{4 \omega_{p;1} \omega_{p;2}}}  e^{i (\omega_{p;2}+\omega_{p;1})t_0} \ .
\end{equation}
Equations \eqref{AuxTensC1} to \eqref{AuxTensC4} become
\begin{eqnarray*}
 L_{\tau \tau} (u_{\pmb{p};j}, u_{\pmb{p};j}) &=& \frac{1}{2(2\pi)^3} \left \lbrace |\chi_{p;j}|^2 \left(p^2 + m_j^2 \right) + |\dot{\chi}_{p;j}|^2 \right \rbrace = \frac{\omega_{p;j}}{2(2\pi)^3} \\
 L_{\tau \tau} (u_{\pmb{p};j}, u^*_{-\pmb{p};j}) &=& \frac{1}{2(2\pi)^3} \left \lbrace (\chi^*_{p;j})^2 \left(p^2 + m_j^2 \right) + (\dot{\chi}^*_{p;j})^2 \right \rbrace = 0 \\
 L_{kk}(u_{\pmb{p};j}, u_{\pmb{p};j}) &=& \frac{1}{2 (2\pi)^3} \left \lbrace |\chi_{p;j}|^2 \left(2p_k^2 - p^2 -m_j^2 \right) +|\dot{\chi}_{p;j}|^2 \right \rbrace = \frac{p_k^2}{2(2\pi)^3 \omega_{p;j}} \\
  L_{kk}(u_{\pmb{p};j}, u^*_{-\pmb{p};j}) &=& \frac{1}{2 (2\pi)^3} \left \lbrace (\chi^*_{p;j})^2 \left(2p_k^2 - p^2 -m_j^2 \right) +(\dot{\chi}^*_{p;j})^2 \right \rbrace = \frac{\left(p_k^2 - \omega_{p;j}^2\right)}{2(2\pi)^3 \omega_{p;j}}e^{2i \omega_{p;j}t} \ .
\end{eqnarray*}
Therefore
\begin{equation}
 \mathbb{T}_{t t}^{(MIX)} = \frac{\sin^2 \theta}{(2\pi)^3} \int d^3 p \frac{(\omega_{p;1}- \omega_{p;2})^2\left(\omega_{p;1} + \omega_{p;2} \right)}{4 \omega_{p;1} \omega_{p;2}} \ .
\end{equation}
The $kk$ component is slightly more involved:
\begin{eqnarray}
 \nonumber \mathbb{T}_{kk}^{(MIX)} &=& \frac{\sin^2 \theta}{(2\pi)^3} \int d^3 p \Bigg \lbrace \frac{\left( \omega_{p;1}- \omega_{p;2}\right)^2}{4 \omega_{p;1}\omega_{p;2}} \sum_{j=1,2} \frac{p_k^2}{\omega_{p;j}} + \left[\frac{\left(\omega_{p;1}^2 - \omega_{p;2}^2 \right) \left( p_{k}^2 - \omega_{p;1}^2\right)}{8 \omega_{p;1}^2 \omega_{p;2}} e^{2 i \omega_{p;1}(t-t_0)} + c.c \right] \\
 &-& \left[\frac{\left(\omega_{p;1}^2 - \omega_{p;2}^2 \right) \left( p_{k}^2 - \omega_{p;2}^2\right)}{8 \omega_{p;1} \omega_{p;2}^2} e^{2 i \omega_{p;2}(t_0-t)} + c.c \right]\Bigg \rbrace \ .
\end{eqnarray}
In particular, when $t_0 = t$,
\begin{eqnarray*}
 \nonumber \mathbb{T}_{kk}^{(MIX)} &=& \frac{\sin^2 \theta}{(2\pi)^3} \int d^3 p \left \lbrace \frac{\left( \omega_{p;1}- \omega_{p;2}\right)^2}{4 \omega_{p;1}\omega_{p;2}} \left(\frac{p_k^2}{\omega_{p;1}} + \frac{p_k^2}{\omega_{p;2}} \right) + \frac{\left(\omega_{p;1}^2 - \omega_{p;2}^2 \right)\left( p_{k}^2 - \omega_{p;1}^2\right)}{4\omega_{p;1}^2 \omega_{p;2}} -\frac{\left(\omega_{p;1}^2 - \omega_{p;2}^2 \right) \left( p_{k}^2 - \omega_{p;2}^2\right)}{4 \omega_{p;1} \omega_{p;2}^2}\right \rbrace \\
 &=& - \frac{\sin^2 \theta}{(2\pi)^3} \int d^3 p \frac{(\omega_{p;1}- \omega_{p;2})^2\left(\omega_{p;1} + \omega_{p;2} \right)}{4 \omega_{p;1} \omega_{p;2}} \ .
\end{eqnarray*}
It follows that
\begin{equation*}
 \rho^{(MIX)} (t,t) = - p^{(MIX)} (t,t) \ .
\end{equation*}
We have recovered the standard result \cite{CapDark1,CapDark2,CapDark3,CapDark4,CapDark5} in flat spacetime, corresponding to the equation of state $w^{(MIX)}(t,t) =\frac{ p^{(MIX)} (t,t)}{\rho^{(MIX)} (t,t)} = -1$.

\section{De Sitter Expansion}

We now move on to a non-trivial application of the formalism above. The general features of $\mathbb{T}_{\mu \nu}^{(MIX)}$ analyzed in the previous section ensure that it may be regarded as the energy-momentum tensor of a perfect fluid. Since it also satisfies the Bianchi identity (see Appendix \ref{BianchiAppendix}), it represents a valid source term for the Einstein field equations, with a metric of the form given by Eq. \eqref{LineElement}. Nevertheless the simultaneous solution of the Einstein field equations with the source term $\mathbb{T}_{\mu \nu}^{(MIX)}$ and of the mode equations
\begin{equation}\label{ReducedModeEquation}
 \ddot{\chi}_{p;j} + \left(p^2 + m^2_j C^2 - \ddot{C} C^{-1} \right)\chi_{p;j} = 0
\end{equation}
is an extremely difficult analyitical task. We postpone this kind of self-consistent computation to future works. Here we instead assume that the Einstein field equations are dominated by some other classical source term $T^{C}_{\mu \nu}$ which forces a definite form of the scale factor $C(t)$ and thus of the metric. As a first approximation we ignore the effect of $\mathbb{T}_{\mu \nu}^{(MIX)}$ on the scale factor, and compute its value on the metric determined by $T^{(C)}_{\mu \nu}$.

Here we take into account a De Sitter expansion, and assume that the scale factor has the form $C(t) = e^{H_0 t}$, for some constant $H_0$ with dimensions of a mass.

\subsection{Mode functions}

For the exponential scale factor $C(t) = e^{H_0 t}$ the conformal time is $\tau = \frac{-e^{-H_0 t}}{H_0} < 0$, and $C(\tau) = -(H_0 \tau)^{-1}$. The mode equations \eqref{ReducedModeEquation} are
\begin{equation}
 \ddot{\chi}_{p;j} + \left(p^2 + \frac{m_j^2}{H_0^2 \tau^2} - \frac{2}{\tau^2} \right) \chi_{p;j} = 0 \ ,
\end{equation}
or, introducing the positive variable $s = -p \tau$
\begin{equation}
 \partial^2_s \chi_{p;j} + \left(1 + \frac{\frac{m_j^2}{H_0^2}-2}{s^2} \right) \chi_{p;j} = 0 \ .
\end{equation}
This is a Bessel-like equation and its general solution can be written as
\begin{equation}
 \chi_{p;j} (s) = s^{\frac{1}{2}} \left( A H^{1}_{\nu_{j}} (s) + B H^{2}_{\nu_j} (s) \right) \ ; \ \ \ \ \ \ \ \nu_j = \sqrt{\frac{9}{4}-\frac{m_j^2}{H_0^2}}
\end{equation}
where $H^{1,2}_{\nu}$ are the Hankel functions of the first and second kind \cite{Abramowitz} while $A,B$ are complex constants. We select the positive energy modes with respect to $\partial_{\tau}$ by requiring that at early times $(\tau \rightarrow - \infty, s \rightarrow \infty)$ the mode functions are proportional to $\chi_{p;j} \propto e^{-ip\tau} = e^{is}$. Given the asymptotic form of the Hankel functions \cite{Abramowitz} for $s \rightarrow \infty$ $H^1_{\nu} \simeq \sqrt{\frac{2}{\pi s}} e^{i \left( s - \frac{\nu \pi}{2} - \frac{\pi}{4}\right)}$ and $H^2_{\nu} \simeq \sqrt{\frac{2}{\pi s}} e^{-i \left( s - \frac{\nu \pi}{2} - \frac{\pi}{4}\right)}$, we set $B=0$. The remaining constant is determined by the normalization condition
\begin{equation}
 i(\chi^*_{p;j} \dot{\chi}_{p;j} - \dot{\chi}^*_{p;j} \chi_{p;j}) = \frac{4p |A|^2}{\pi} e^{\pi \Im(\nu_j)} \overset{!}{=} 1
\end{equation}
which implies $A=e^{- \frac{\pi \Im (\nu_j)}{2}} \sqrt{\frac{\pi}{4 p}}$ up to a phase. The modes then take the form
\begin{equation}\label{ModeSolution}
 \chi_{p;j} (\tau) = e^{-\frac{\pi \Im(\nu_j)}{2}} \sqrt{- \frac{\pi \tau}{4}}H^1_{\nu_j} (-p\tau) \ .
\end{equation}
The mode index $\nu_j$ turns out to be imaginary, at least in relatively close epochs, since $\frac{m_j}{H_0} \gg \frac{3}{2}$. To give an idea of the size of $H_0$, consider that the current Hubble constant is of the order $H_0 \sim 10^{-33} \mathrm{eV}$, which is far below the masses of the known particles. Even for very light masses, say $m_j \simeq 10^{-2} \mathrm{eV}$, we can expect the condition $\frac{m_j}{H_0} \gg \frac{3}{2}$ to break down only very close to the Big Bang, when $H_0$ blows up due to inflation. Then $\nu_j = i |\nu_j| = i \sqrt{\frac{m_j^2}{H_0^2}-\frac{9}{4}}$. We shall later need the asymptotic form of Eq. \eqref{ModeSolution} at late times $\tau \rightarrow 0^{-}, s \rightarrow 0$. Because $\Re (\nu_j)= 0$, we cannot directly use Eq. (9.1.9) of \cite{Abramowitz} and have first to express the Hankel functions in terms of Bessel functions. We quote the resulting expression for $H^1_{\nu}$
\begin{equation}\label{AsymptoticModes}
 H^{1}_{\nu_j} (-p \tau \rightarrow 0) \simeq \frac{1}{\sinh \pi |\nu_j|} \left[\frac{e^{\pi |\nu_j|}}{\Gamma(1+\nu_j)} \left(\frac{-p\tau}{2} \right)^{\nu_j} - \frac{1}{\Gamma(1-\nu_j)} \left(\frac{-p\tau}{2} \right)^{-\nu_j} \right]
\end{equation}
where $\Gamma (x)$ is the Euler Gamma function. We can immediately write down the components of the auxiliary tensor Eqs. \eqref{AuxTensC1} to \eqref{AuxTensC4} as
\begin{eqnarray}\label{AuxTensDs1}
 \nonumber L_{\tau \tau} (u_{\pmb{p};j}, u_{\pmb{p};j}) &=& \frac{\pi H_0^2 \tau^2 e^{- \pi |\nu_j|}}{8 (2 \pi)^3} \left \lbrace |H^1_{\nu_j}|^2 \left(- p^2 \tau - \frac{ m_j^2}{ H_0^2 \tau} - \frac{1}{4 \tau} \right) -  \tau|\partial_{\tau}H^1_{\nu_j}|^2 - \frac{1}{2} \left( H^{1*}_{\nu_j} \partial_{\tau} H^{1}_{\nu_j} + H^{1}_{\nu_j}\partial_{\tau}H^{1*}_{\nu_j}\right)\right \rbrace \\
 \nonumber L_{\tau \tau} (u_{\pmb{p};j}, u^*_{-\pmb{p};j}) &=& \frac{\pi H_0^2 \tau^2 e^{- \pi |\nu_j|}}{8 (2 \pi)^3} \left \lbrace (H^{1*}_{\nu_j})^2 \left(- p^2 \tau - \frac{ m_j^2}{ H_0^2 \tau} - \frac{1}{4 \tau} \right) -  \tau(\partial_{\tau}H^{1*}_{\nu_j})^2 -  \left( H^{1*}_{\nu_j} \partial_{\tau} H^{1*}_{\nu_j}\right)\right \rbrace \\
 \nonumber L_{k k} (u_{\pmb{p};j}, u_{\pmb{p};j}) &=& \frac{\pi H_0^2 \tau^2 e^{- \pi |\nu_j|}}{8 (2 \pi)^3} \left \lbrace |H^1_{\nu_j}|^2 \left(- 2p_k^2 \tau + p^2 \tau + \frac{ m_j^2}{ H_0^2 \tau} - \frac{9}{4 \tau} \right) -  \tau|\partial_{\tau}H^1_{\nu_j}|^2 - \frac{3}{2} \left( H^{1*}_{\nu_j} \partial_{\tau} H^{1}_{\nu_j} + H^{1}_{\nu_j}\partial_{\tau}H^{1*}_{\nu_j}\right)\right \rbrace \\
  L_{k k} (u_{\pmb{p};j}, u^*_{-\pmb{p};j}) &=& \frac{\pi H_0^2 \tau^2 e^{- \pi |\nu_j|}}{8 (2 \pi)^3} \left \lbrace (H^{1*}_{\nu_j})^2 \left(-2p_k^2 \tau + p^2 \tau + \frac{ m_j^2}{ H_0^2 \tau} - \frac{9}{4 \tau} \right) -  \tau(\partial_{\tau}H^{1*}_{\nu_j})^2 -  3\left( H^{1*}_{\nu_j} \partial_{\tau} H^{1*}_{\nu_j}\right)\right \rbrace \ .
\end{eqnarray}
Here we have suppressed the argument $-p \tau$ of the Hankel functions. Likewise we can compute the Bogoliubov coefficients of flavor mixing by inserting the solutions of Eq. \eqref{ModeSolution} in the definition of Eq. \eqref{BogoliubovCoefficients2}:
\begin{eqnarray}\label{BogoliubovCoefficientsDS1}
 \nonumber \Lambda_p (\tau) &=& i e^{-\frac{\pi (|\nu_2| + |\nu_1|)}{2}}\left(\frac{-\pi \tau}{4}\right) \left(H^{1*}_{\nu_2} (-p\tau) \partial_{\tau} H^{1}_{\nu_1} (-p\tau) - H^{1}_{\nu_1}(-p\tau) \partial_{\tau} H^{1*}_{\nu_2} (-p\tau)\right) \\
 \Xi_p (\tau) &=& i e^{-\frac{\pi (|\nu_2| + |\nu_1|)}{2}}\left(\frac{-\pi \tau}{4}\right) \left(H^{1*}_{\nu_1} (-p\tau) \partial_{\tau} H^{1*}_{\nu_2} (-p\tau) - H^{1*}_{\nu_2}(-p\tau) \partial_{\tau} H^{1*}_{\nu_1} (-p\tau)\right)
\end{eqnarray}

\subsection{The late time approximation}

Above we have obtained fairly simple analytical expressions for the various quantities that appear in the energy-momentum tensor $\mathbb{T}^{(MIX)}_{\mu \nu}$. The exact value of $\rho^{(MIX)}$ and $p^{(MIX)}$ for given masses and conformal time can be obtained by inserting the mixing Bogoliubiov coefficients of Eq. \eqref{BogoliubovCoefficientsDS1} and the auxiliary tensor of Eq. \eqref{AuxTensDs1} in Eq. \eqref{ReducedEnergyMomentumTensor} for the corresponding components. At this point it is impossible to proceed analytically in the computation of the final $\int d^3 p$ integral, which involves a non trivial combination of $p$-dependent Hankel functions. The integral may be evaluated numerically for given sets of parameters.

We prefer here to pursue a different route, and gain insight on $\rho^{(MIX)}$ and $p^{(MIX)}$ by invoking some approximations and handling the approximate integral analytically. We focus on the late time approximation, in which both the flavor vacuum and the energy-momentum tensor operator are considered for late conformal time arguments, $\tau_0 \rightarrow 0^{-}$ and $\tau \rightarrow 0^{-}$ respectively. As we wish to compute the energy density and the pressure at a given time $\tau$ associated to the flavor vacuum defined at a previous (or at most coincident) time, we shall always consider $\tau \geq \tau_0$. The late time approximation has the advantage to turn the Hankel functions integral into an integral over  polynomials in $p$, which can be straightforwardly evaluated.
With a tedious but simple computation we determine the late time form of $\mathbb{T}^{(MIX)}_{\tau \tau}$ and $\mathbb{T}^{(MIX)}_{kk}$. The computation employs the asymptotic form of the Hankel functions of Eq. \eqref{AsymptoticModes} and makes use of the properties of the $\Gamma$ functions in several intermediate steps. We show the result in the Appendix  \ref{EnMom}. The expressions derived in Eqs. \eqref{MassIntegral}, \eqref{EnergyIntegral} and \eqref{PressureIntegral} are still quite cumbersome to deal with. Yet we can invoke another approximation regarding the size of the mass parameters $m_1,m_2$. As argued above, except for epochs extremely close to the Big Bang, we can expect that even very small masses $ \sim 10^{-2} \mathrm{eV}$ shall be several orders of magnitude greater than the expansion rate $H_0$. Then we can safely consider the high mass limit
\begin{equation}
 \frac{m_j}{H_0} \gg 1 \Longrightarrow|\nu_j| \gg 1 \ ,
\end{equation}
and take into account the $|\nu_j| \rightarrow \infty$ asymptotic expression for each of Eqs. \eqref{MassIntegral}, \eqref{EnergyIntegral} and \eqref{PressureIntegral}. In such a limit many of the terms are suppressed by hyperbolic functions $\sinh( \pi |\nu_j|)$ in the denominator and only a few terms survive. We find

\begin{eqnarray}\label{EnergyDensityDS}
 \nonumber  \mathbb{T}^{(MIX)}_{\tau \tau} &\simeq& \frac{\sin^2 \theta H_0^2 \tau}{(2\pi)^3} \int d^3 p \Bigg \lbrace \left(\frac{(|\nu_1|^2 + |\nu_2|^2)\coth(\pi |\nu_1|)\coth (\pi |\nu_2|)}{2|\nu_1||\nu_2|} - 1\right) \sum_j \frac{\coth(\pi |\nu_j|)}{|\nu_j|}\left(\frac{1}{2} - \frac{m_j^2}{2H_0^2} \right) + \\
\nonumber &-& \frac{5(|\nu_2|^2 -|\nu_1|^2)\coth (\pi |\nu_1|)}{32|\nu_1||\nu_2|^2} \left(1 + \coth^2 (\pi |\nu_2|)  \right) \cos \left(2 |\nu_2| \log \left(\frac{\tau}{\tau_0} \right) \right) + \\
\nonumber &-& \frac{5(|\nu_1|^2 -|\nu_2|^2)\coth (\pi |\nu_2|)}{32|\nu_1|^2|\nu_2|} \left(1 + \coth^2 (\pi |\nu_1|)  \right) \cos \left(2 |\nu_1| \log \left(\frac{\tau}{\tau_0} \right) \right) + \\
\nonumber &+& \frac{\coth(\pi |\nu_1|) (|\nu_2|^2 -|\nu_1|^2)}{16|\nu_1||\nu_2|}\left(1 + \coth^2 (\pi |\nu_2|) \right) \sin \left(2 |\nu_2| \log \left(\frac{\tau}{\tau_0}\right)\right) + \\
&+& \frac{\coth(\pi |\nu_2|) (|\nu_1|^2 -|\nu_2|^2)}{16|\nu_1||\nu_2|}\left(1 + \coth^2 (\pi |\nu_1|) \right) \sin \left(2 |\nu_1| \log \left(\frac{\tau}{\tau_0}\right)\right) \Bigg \rbrace \ .
\end{eqnarray}

and

\begin{eqnarray}\label{PressureDS}
 \nonumber \mathbb{T}^{(MIX)}_{kk} &\simeq& \frac{\sin^2 \theta H_0^2 \tau}{(2\pi)^3} \int d^3 p \Bigg \lbrace \frac{\coth(\pi|\nu_1|) (|\nu_2|^2 -|\nu_1|^2)}{8|\nu_1|}\left( 1 + \coth^2 (\pi |\nu_2|)\right) \cos \left(2 |\nu_2| \log \left(\frac{\tau}{\tau_0} \right) \right) + \\
 \nonumber &+& \frac{\coth(\pi|\nu_2|) (|\nu_1|^2 -|\nu_2|^2)}{8|\nu_2|}\left( 1 + \coth^2 (\pi |\nu_1|)\right) \cos \left(2 |\nu_1| \log \left(\frac{\tau}{\tau_0} \right) \right) \\
 \nonumber &+& \frac{3 \coth(\pi |\nu_1|)(|\nu_2|^2 - |\nu_1|^2)}{16 |\nu_1| |\nu_2|} \left(1 + \coth^2 (\pi |\nu_2|) \right) \sin \left(2 |\nu_2| \log \left(\frac{\tau}{\tau_0} \right) \right) \\
&+& \frac{3 \coth(\pi |\nu_2|)(|\nu_1|^2 - |\nu_2|^2)}{16 |\nu_1| |\nu_2|} \left(1 + \coth^2 (\pi |\nu_1|) \right) \sin \left(2 |\nu_1| \log \left(\frac{\tau}{\tau_0} \right) \right) \Bigg \rbrace \ .
\end{eqnarray}
We can see from the above equations that both the $\tau \tau$ component and the $k k$ component depend only on the ratio $\frac{\tau}{\tau_0}$ between the instantaneous $\tau$ and the reference conformal time $\tau_0$. They exhibit a simple oscillating behaviour weighted by the mode indices $\nu_j$. The integral over $d^3 p$ is trivial but divergent. In order to obtain a finite result we need a regularization. Given that the particle mixing phenomena are generally suppressed at high momenta, that is, when the mass terms that are responsible for mixing become negligible, we employ an ultraviolet cutoff $\mathcal{P}_0$. For instance, if the bosons which are being mixed are the supersymmetric partners of neutrinos, the natural choice is  a cutoff of the order of the electroweak scale $\mathcal{P}_0 \simeq 246 \ \mathrm{GeV}$. We must now recall that the mode label $\pmb{p}$ is \emph{not} the actual momentum carried by the particles, which is instead the comoving momentum $\pmb{p}_{PHYS} = \frac{\pmb{p}}{C(\tau)}$. If the cutoff is to be imposed on the physical momentum of the particles $p_{PHYS}^{CUTOFF} = \mathcal{P}_0$, then the cutoff for the mode label is the comoving cutoff
\begin{equation}
 p^{CUTOFF} = p_{PHYS}^{CUTOFF} C(\tau) = -\frac{\mathcal{P}_0}{H_0\tau} \equiv \mathcal{P}(\tau) \ ,
\end{equation}
where of course, given that $\tau < 0$, $\mathcal{P}(\tau)$ is strictly positive. Moving to polar coordinates, evaluating the angular integral and performing the $dp$ integral with the cutoff $\mathcal{P} (\tau)$, we find

\begin{eqnarray}\label{EnergyDensityDS2}
 \nonumber  \mathbb{T}^{(MIX)}_{\tau \tau} &\simeq& \frac{\sin^2 \theta H_0^2 \tau \mathcal{P}^3 (\tau)}{6 \pi^2}  \Bigg \lbrace \left(\frac{(|\nu_1|^2 + |\nu_2|^2)\coth(\pi |\nu_1|)\coth (\pi |\nu_2|)}{2|\nu_1||\nu_2|} - 1\right) \sum_j \frac{\coth(\pi |\nu_j|)}{|\nu_j|}\left(\frac{1}{2} - \frac{m_j^2}{2H_0^2} \right) + \\
\nonumber &-& \frac{5(|\nu_2|^2 -|\nu_1|^2)\coth (\pi |\nu_1|)}{32|\nu_1||\nu_2|^2} \left(1 + \coth^2 (\pi |\nu_2|)  \right) \cos \left(2 |\nu_2| \log \left(\frac{\tau}{\tau_0} \right) \right) + \\
\nonumber &-& \frac{5(|\nu_1|^2 -|\nu_2|^2)\coth (\pi |\nu_2|)}{32|\nu_1|^2|\nu_2|} \left(1 + \coth^2 (\pi |\nu_1|)  \right) \cos \left(2 |\nu_1| \log \left(\frac{\tau}{\tau_0} \right) \right) + \\
\nonumber &+& \frac{\coth(\pi |\nu_1|) (|\nu_2|^2 -|\nu_1|^2)}{16|\nu_1||\nu_2|}\left(1 + \coth^2 (\pi |\nu_2|) \right) \sin \left(2 |\nu_2| \log \left(\frac{\tau}{\tau_0}\right)\right) + \\
&+& \frac{\coth(\pi |\nu_2|) (|\nu_1|^2 -|\nu_2|^2)}{16|\nu_1||\nu_2|}\left(1 + \coth^2 (\pi |\nu_1|) \right) \sin \left(2 |\nu_1| \log \left(\frac{\tau}{\tau_0}\right)\right) \Bigg \rbrace \ .
\end{eqnarray}

and

\begin{eqnarray}\label{PressureDS2}
 \nonumber \mathbb{T}^{(MIX)}_{kk} &\simeq& \frac{\sin^2 \theta H_0^2 \tau \mathcal{P}^3 (\tau)}{6 \pi^2} \Bigg \lbrace \frac{\coth(\pi|\nu_1|) (|\nu_2|^2 -|\nu_1|^2)}{8|\nu_1|}\left( 1 + \coth^2 (\pi |\nu_2|)\right) \cos \left(2 |\nu_2| \log \left(\frac{\tau}{\tau_0} \right) \right) + \\
 \nonumber &+& \frac{\coth(\pi|\nu_2|) (|\nu_1|^2 -|\nu_2|^2)}{8|\nu_2|}\left( 1 + \coth^2 (\pi |\nu_1|)\right) \cos \left(2 |\nu_1| \log \left(\frac{\tau}{\tau_0} \right) \right) \\
 \nonumber &+& \frac{3 \coth(\pi |\nu_1|)(|\nu_2|^2 - |\nu_1|^2)}{16 |\nu_1| |\nu_2|} \left(1 + \coth^2 (\pi |\nu_2|) \right) \sin \left(2 |\nu_2| \log \left(\frac{\tau}{\tau_0} \right) \right) \\
&+& \frac{3 \coth(\pi |\nu_2|)(|\nu_1|^2 - |\nu_2|^2)}{16 |\nu_1| |\nu_2|} \left(1 + \coth^2 (\pi |\nu_1|) \right) \sin \left(2 |\nu_1| \log \left(\frac{\tau}{\tau_0} \right) \right) \Bigg \rbrace \ .
\end{eqnarray}
Although the energy density and the pressure (which differ from Eqs. \eqref{EnergyDensityDS2} and \eqref{PressureDS2} by the multiplicative factor $C^{-2}(\tau) = H_0^2 \tau^2$) depend on the cutoff, their ratio is cutoff-independent. Then we can derive a cutoff-indipendent equation of state dividing Eq. \eqref{PressureDS2} by \eqref{EnergyDensityDS2}

\begin{equation}
 w^{(MIX)} (\tau_0, \tau) = \frac{p^{(MIX)}(\tau_0,\tau)}{\rho^{(MIX)}(\tau_0,\tau)} = \frac{\mathbb{T}^{(MIX)}_{kk}(\tau_0,\tau)}{\mathbb{T}^{(MIX)}_{\tau \tau}(\tau_0,\tau)} \ .
\end{equation}

\begin{figure}
\centering
\includegraphics[width=0.48\linewidth]{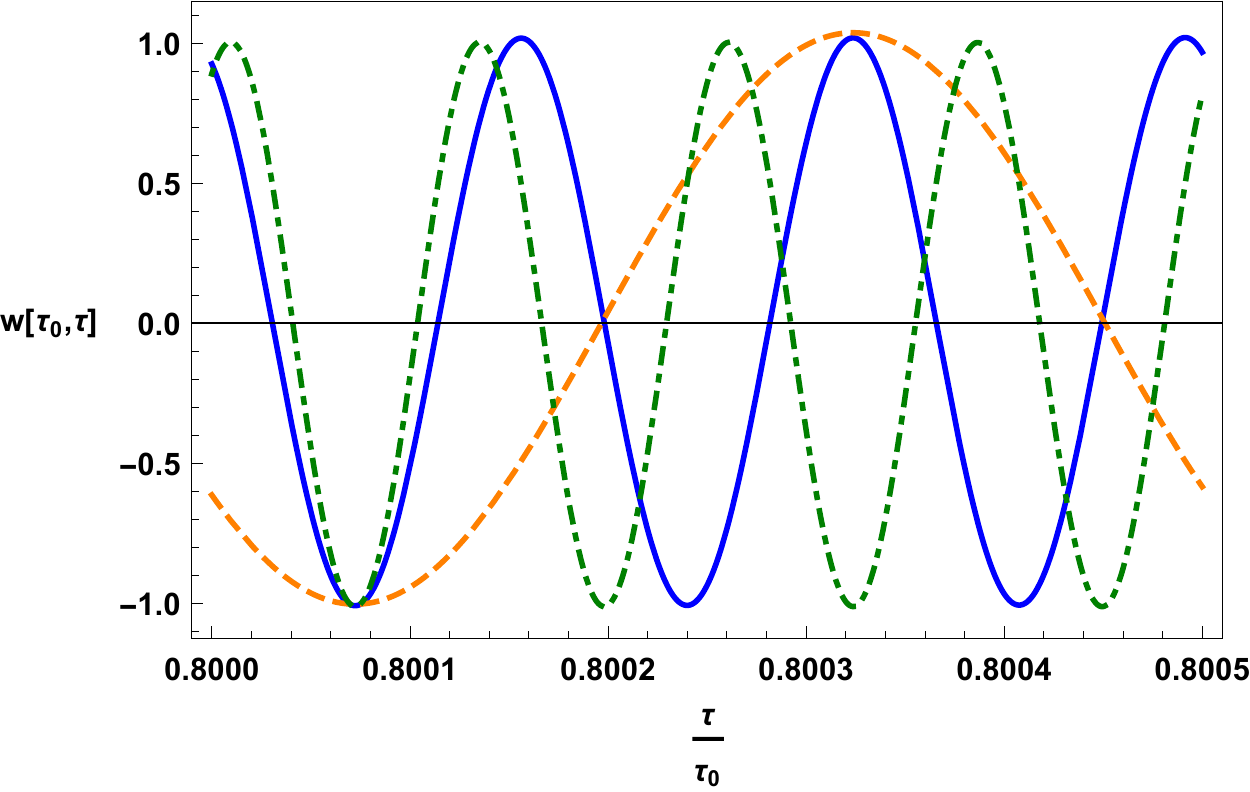}
\includegraphics[width=0.48\linewidth]{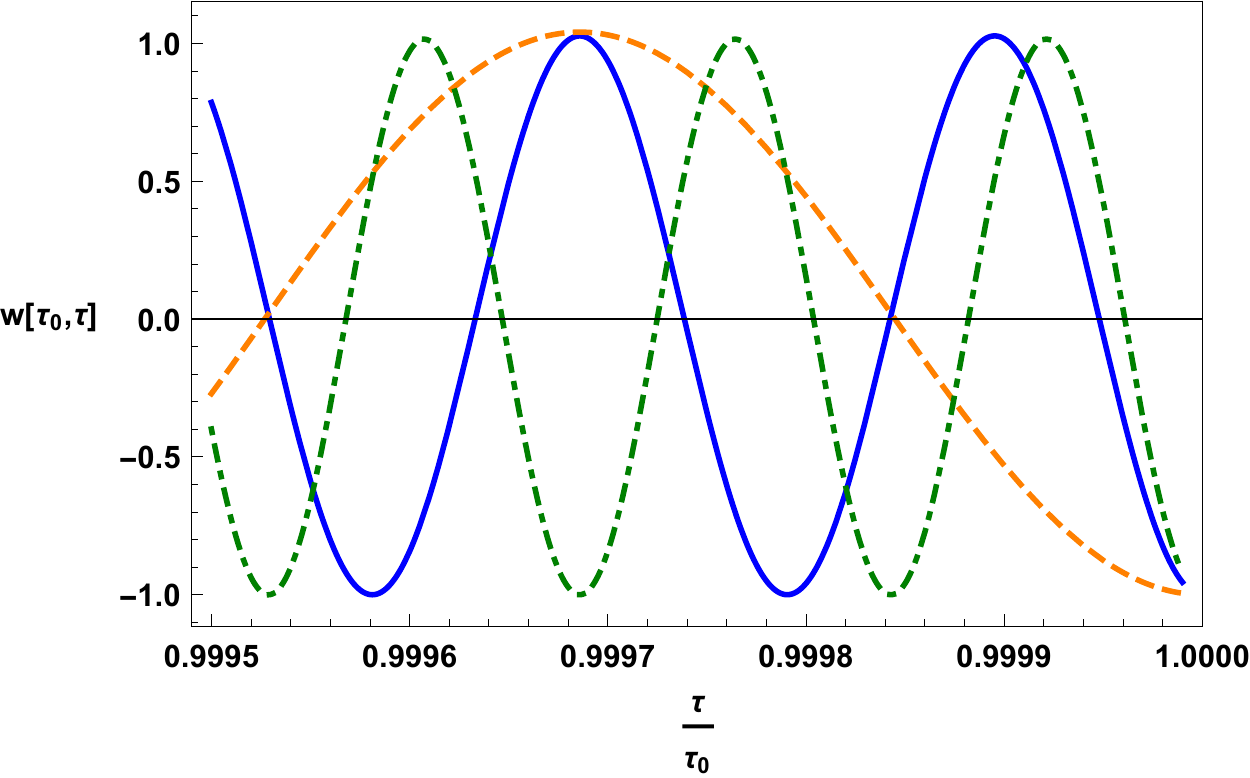}
\caption{
(color online) Plots of the adiabatic factor $w^{(MIX)} ( \tau_0,\tau)$ for sample values of the parameters. Masses are expressed in units of $H_0$ and conformal times in units of $H_0^{-1}$. Masses are chosen so to satisfy the condition $m_j \gg H_0$. Blue solid line: $m_1 = 200, m_2 = 15000$; orange dashed line: $m_1 = 100, m_2 = 5000$; dark green dotdashed line: $m_1 = 150, m_2 = 20000$.
}
\label{EoSPlot}
\end{figure}

The behaviour of $w^{(MIX)}$ is shown in Fig. 1 for sample values of the parameters. Interestingly, the adiabatic factor undergoes oscillations in the full range $[-1,1]$, periodically going through ``dark energy'' phases ($w < -\frac{1}{3}$), dust phases ($w=0$) and radiation phases ($w=\frac{1}{3}$). This is at odds with both the fermionic counterpart ($w=0$ at all times) and the flat spacetime result ($w=-1$ at all times). From the right panel of Fig. 1 one can see that all the curves approach the limiting value $w=-1$ when $\tau \rightarrow \tau_0$ and their ratio tends to unity. This is nothing but a reinstatement of the flat space limit: the instantaneous adiabatic factor is the same for any choice of parameters and is equal to

\begin{equation}
 w^{(MIX)}(\tau,\tau) = -1 \ .
\end{equation}

\begin{figure}
\centering
\includegraphics[width=0.76\linewidth]{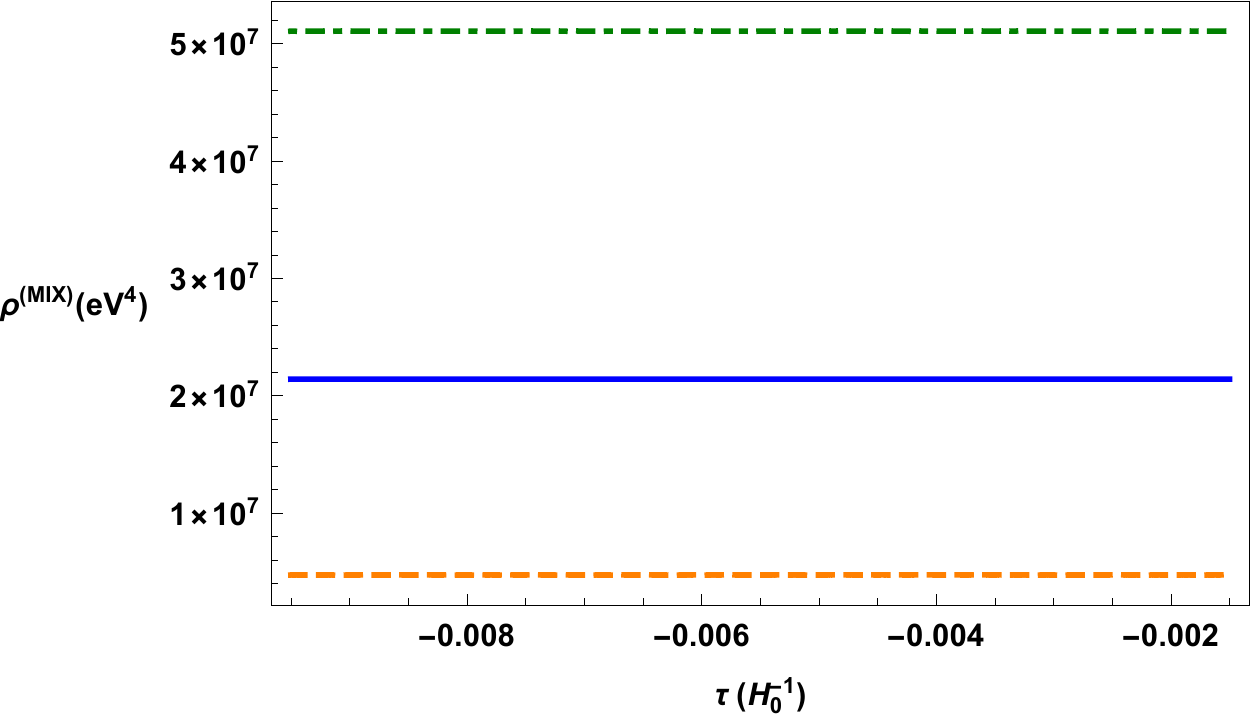}
\caption{
(color online) Plots of the energy density $\rho^{(MIX)} ( \tau_0,\tau)$ for sample values of the parameters. Masses are expressed in units of $H_0$ and conformal times in units of $H_0^{-1}$. We have considered $\sin^2 \theta = 0.307$, $\mathcal{P}_0=246 \mathrm{GeV}$, $H_0 = 10^{-33} \mathrm{eV}$ and $\tau_0 = -0.01$. Masses are chosen so to satisfy the condition $m_j \gg H_0$ and conformal times to suite the late time approximation. Blue solid line: $m_1 = 200, m_2 = 15000$; orange dashed line: $m_1 = 100, m_2 = 5000$; dark green dotdashed line: $m_1 = 150, m_2 = 20000$.
}
\label{EDenPlot}
\end{figure}
The behaviour of the energy density $\rho^{(MIX)} (\tau_0, \tau) $ for sample values of the parameters is shown in Fig. 2. We can see that within the approximations employed, the energy density is a constant for fixed values of the parameters.

\section{Conclusions}

We have analyzed the possible contribution  to the dark energy given by  the quantum vacuum for mixed boson, named flavor vacuum,  by studying   the boson mixing in curved space.
We have calculated  the expectation value of the energy momentum tensor $T_{\mu \nu}^{MIX}$ of free bosons on the vacuum for mixed fields for a spatially flat FLRW metric, and  we have shown that this tensor is diagonal (the off-diagonal elements are all equal to zero), and therefore, it behaves as  a classical perfect fluid, whose properties depend on the space-time geometry.
This result connects the non trivial structure of the quantum flavor vacuum for mixed bosons with  classical fluids.
Moreover,   $T_{\mu \nu}^{MIX}$  also represents a valid source term for the Einstein field equations, since it
satisfies the Bianchi identity. However, as approximation, we ignore the effect of $T_{\mu \nu}^{MIX}$
  on the scale factor, and compute its value on the metric determined only by  classical source terms.

We have analyzed the trend of $T_{\mu \nu}^{MIX}$  in the case of de Sitter background, and we have shown that the values of the adiabatic factor of the boson flavor mixing $w^{(MIX)}$  are included in the interval $[-1;  1]$. On the other hand, in time limit of the flat space time, the bosonic mixed vacuum assumes the state equation of the dark energy: $w^{(MIX)}=-1$. Therefore, there may be a strong link between purely quantum effects and phenomena on a cosmological scale, and the vacuum energy of mixed bosons, such as neutrino super-partners,  may give  a non-trivial contribution to the dark energy.

\section*{Acknowledgements}
A.C. and A.Q. acknowledge partial financial support from MUR and INFN, A.C. also acknowledges the COST Action CA1511 Cosmology
and Astrophysics Network for Theoretical Advances and Training
Actions (CANTATA).

\appendix
\section{Bianchi identity}
\label{BianchiAppendix}
In this appendix we prove explicitly the covariant conservation of the energy momentum tensor associated to the flavor vacuum. We demonstrate the four equations
\begin{equation}\label{Divergence}
 \nabla_{\mu} \mathbb{T}^{\mu \nu} = 0
\end{equation}
with $\nabla_{\mu}$ denoting the covariant derivative. There is no need here to distinguish between $\mathbb{T}_{\mu \nu}^{(MIX)}$ and $\mathbb{T}_{\mu \nu}^{(N)}$, since both satisfy eq. \eqref{Divergence}, and so does the full energy-momentum tensor. Let us first compute the connection coefficients for the metric of eq. \eqref{LineElement}. The non-zero coefficients are
\begin{equation}\label{Christoffel}
 \Gamma^{\tau}_{\tau \tau} = \Gamma^{\tau}_{ii} = \Gamma^{i}_{\tau i} = \Gamma^{i}_{i \tau} = \frac{\dot{C}}{C} \ .
\end{equation}
No sum is here intended over repeated indices. Notice that the coefficients depend only on $\tau$. In terms of the connection coefficients, the covariant divergence reads
\begin{equation}\label{ExplicitDivergence}
 \nabla_{\mu} \mathbb{T}^{\mu \nu} = \partial_{\mu} \mathbb{T}^{\mu \nu} + \Gamma^{\mu}_{\mu \sigma} \mathbb{T}^{\sigma \nu} + \Gamma^{\nu}_{\mu \sigma} \mathbb{T}^{\mu \sigma} \ .
\end{equation}
\begin{itemize}
 \item ($\nu = i$) For $\nu = i$, with $i=1,2,3$ equation \eqref{ExplicitDivergence} becomes
 \begin{equation}
  \nabla_{\mu}\mathbb{T}^{\mu i} = \partial_{\mu} \mathbb{T}^{\mu i} + \Gamma^{\mu}_{\mu \sigma} \mathbb{T}^{\sigma i} + \Gamma^{i}_{\mu \sigma} \mathbb{T}^{\mu \sigma} \ .
 \end{equation}
From the diagonality of $\mathbb{T}^{\mu \nu}$ proved above, we can write
 \begin{equation}\label{Divergence1}
  \nabla_{\mu}\mathbb{T}^{\mu i} =  \partial_{i} \mathbb{T}^{i i} + \sum_{\mu} \Gamma^{\mu}_{\mu i} \mathbb{T}^{i i} + \sum_{\mu} \Gamma^{i}_{\mu \mu} \mathbb{T}^{\mu \mu} \ ,
 \end{equation}
 where no sum is intended over repeated indices and the summations are written out explicitly to avoid confusion. The first term on the right hand side of eq. \eqref{Divergence1} is zero, since $\mathbb{T}^{\mu \nu}$ depends only on $\tau$. Similarly, from eq. \eqref{Christoffel} we know that $\Gamma^{\mu}_{\mu i} = 0 = \Gamma^{i}_{\mu \mu}$ for each $\mu = 0,1,2,3$ and each $i=1,2,3$, so that also the second and the third term on the right hand side of eq. \eqref{Divergence1} vanish. Therefore
 \begin{equation}
  \nabla_{\mu} \mathbb{T}^{\mu i} = 0 \ \ \ \ \forall i \ .
 \end{equation}
\item ($\nu = \tau$) Only a slightly longer calculation is needed to prove the statement for $\nu = \tau$. Starting from equation \eqref{ExplicitDivergence} we have
\begin{eqnarray}\label{Divergence2}
 \nonumber \nabla_{\mu} \mathbb{T}^{\mu \tau} &=& \partial_{\mu}\mathbb{T}^{\mu \tau} + \Gamma^{\mu}_{\mu \sigma} \Gamma^{\sigma \tau} + \Gamma^{\tau}_{\mu \sigma} \mathbb{T}^{\mu \sigma} \\
 \nonumber &=& \partial_{\tau} \mathbb{T}^{\tau \tau} + \left(\Gamma^{\tau}_{\tau \tau} + \sum_i \Gamma^{i}_{i \tau}\right) \mathbb{T}^{\tau \tau} + \Gamma^{\tau}_{\tau \tau} \mathbb{T}^{\tau \tau} + \sum_{i} \Gamma^{\tau}_{ii} \mathbb{T}^{ii} \\ &=& \partial_{\tau} \mathbb{T}^{\tau \tau} + 5 \Gamma^{\tau}_{\tau \tau} \mathbb{T}^{\tau \tau} + 3 \Gamma^{\tau}_{\tau \tau}\mathbb{T}^{ii} \ ,
\end{eqnarray}
where we have used the diagonality and isotropy of $\mathbb{T}^{\mu \nu}$ and eqs. \eqref{Christoffel}. It is convenient to express Eq. \eqref{Divergence2} in terms of the covariant components by means of the metric of Eq. \eqref{LineElement}, so to get
\begin{equation}\label{Divergence3}
 \nabla_{\mu} \mathbb{T}^{\mu \tau} = \partial_{\tau} \left(C^{-4}\mathbb{T}_{\tau \tau} \right) + 5 C^{-5} \dot{C} + 3 C^{-5} \dot{C} \mathbb{T}_{ii} \ .
\end{equation}
From equation \eqref{VeV}, we know that each of the terms above is the integral of the auxiliary tensor components $L_{\tau \tau}, L_{i i}$ weighted by $\tau$-independent coefficients (because the Bogoliubov coefficients are evaluated at the fixed reference time $\tau_0$). It is then sufficient to prove that
\begin{equation}\label{Div}
  \partial_{\tau} \left( C^{-4} L_{\tau \tau}(A,B) \right) + 5 C^{-5} \dot{C} L_{\tau \tau} (A,B) + 3C^{-5} \dot{C} L_{i i} (A,B) = 0
\end{equation}
for each $A,B = u_{\pmb{p};j}, u^*_{-\pmb{p};j}$, to show that the divergence \eqref{Divergence3} vanishes. It is understood that the equality \eqref{Div} has to hold, multiplied by the appropriate Bogoliubov coefficients, under the integral sign $\int d^3 p$. We can prove Eq. \eqref{Div} by direct computation, inserting Eqs. \eqref{AuxTensC1} to \eqref{AuxTensC4} and using Eq. \eqref{ReducedModeEquation}. We have

\begin{eqnarray*}
&& \partial_{\tau} \left( C^{-4} L_{\tau \tau}(u_{\pmb{p};j},u_{\pmb{p;j}}) \right) + 5 C^{-5} \dot{C} L_{\tau \tau} (u_{\pmb{p};j},u_{\pmb{p};j}) + 3 C^{-5} \dot{C} L_{i i} (u_{\pmb{p},j},u_{\pmb{p};j}) = \\
&& C^{-4} \left \lbrace \partial_{\tau} \left( L_{\tau \tau}(u_{\pmb{p};j},u_{\pmb{p;j}}) \right) + \dot{C}C^{-1}  L_{\tau \tau}(u_{\pmb{p};j},u_{\pmb{p;j}}) +3\dot{C}C^{-1}  L_{i i}(u_{\pmb{p};j},u_{\pmb{p;j}})\right \rbrace = \\
&& \frac{C^{-4}}{2(2\pi)^3} \Bigg \lbrace \left(\dot{\chi}^*_{p;j} \chi_{p;j} + \chi^*_{p;j} \dot{\chi}_{p;j} \right) \left(p^2 C^{-2} + m_j^2 + 4C^{-4}\dot{C}^2 -C^{-3}\ddot{C} \right) + |\chi_{p;j}|^2 \left(- 2p^2 C^{-3} \dot{C} - 4 C^{-5} \dot{C}^3 + 2 C^{-4} \dot{C} \ddot{C} \right) \\
&& + |\dot{\chi}_{p;j}|^2 \left(- 4 C^{-3}\dot{C} \right) + \left(\chi^*_{p;j} \ddot{\chi}_{p;j} + \ddot{\chi}^*_{p;j} \chi_{p;j}\right) \left(- C^{-3} \dot{C} \right) + \left( \ddot{\chi}^*_{p;j} \dot{\chi}_{p;j} + \dot{\chi}^*_{p;j} \ddot{\chi}_{p;j} \right)C^{-2} \Bigg \rbrace \\
&& + \frac{C^{-4}}{2(2\pi)^3} \Bigg \lbrace |\chi_{p;j}|^2 \left(p^2 C^{-3}\dot{C} + m_j^2 C^{-1}\dot{C} + C^{-5}\dot{C}^3 \right) + |\dot{\chi}_{p;j}|^2 C^{-3} \dot{C} - C^{-4}\dot{C}^2 \left(\chi^*_{p;j} \dot{\chi}_{p;j} + \dot{\chi}^*_{p;j} \chi_{p;j} \right) \Bigg \rbrace \\
&& + \frac{3C^{-4}}{2(2\pi)^3} \Bigg \lbrace |\chi_{p;j}|^2 \left(2p_i^2C^{-3}\dot{C} -p^2 C^{-3}\dot{C} - m_j^2 C^{-1}\dot{C} + C^{-5}\dot{C}^3 \right) + |\dot{\chi}_{p;j}|^2 C^{-3} \dot{C} - C^{-4}\dot{C}^2 \left(\chi^*_{p;j} \dot{\chi}_{p;j} + \dot{\chi}^*_{p;j} \chi_{p;j} \right) \Bigg \rbrace = \\
&& \frac{C^{-4}}{2(2\pi)^3} \Bigg \lbrace |\chi_{p;j}|^2 \left(-4p^2 C^{-3} \dot{C} -2 m_j^2 C^{-1}\dot{C} + 6 p_i^2 C^{-3} \dot{C} + 2 C^{-4}\dot{C}\ddot{C} \right) + \left(\chi_{p;j}^* \dot{\chi}_{p;j} + \dot{\chi}^*_{p;j} \chi_{p;j} \right) \left(p^2C^{-2} + m_j^2 -C^{-3}\ddot{C} \right) \\
&&+ \left(\chi^*_{p;j} \ddot{\chi}_{p;j} + \ddot{\chi}^*_{p;j} \chi_{p;j}\right) \left(- C^{-3} \dot{C} \right) + \left( \ddot{\chi}^*_{p;j} \dot{\chi}_{p;j} + \dot{\chi}^*_{p;j} \ddot{\chi}_{p;j} \right)C^{-2} \Bigg \rbrace =
 \frac{C^{-7}\dot{C}}{2(2\pi)^3} \left[ |\chi_{p;j}|^2 \left(6p_i^2 - 2p^2 \right) \right] \ .
\end{eqnarray*}
In the last equality we have made use of Eq. \eqref{ReducedModeEquation} and its complex conjugate. Recalling that this quantity multiplies a function of $p^2$ under the integral sign $\int d^3 p$, we can make the substition\footnote{This is obvious. Given any function $F(p^2)$, one has
\begin{equation*}
 \int d^3 p \ p_1^2 F(p^2) = \int d^3 p \ p_2^2 F(p^2) = \int d^3  p \ p_3^2 F(p^2) \ .
\end{equation*}
Summing one gets $\int d^3 p (p_1^2 + p_2^2 + p_3^2) F(p) = 3 \int d^3 p \ p_k^2 F(p^2) = \int d^3 p \ p^2 F(p^2)$ or
\begin{equation}
 \int d^3 p \ p_k^2 F(p^2) = \int d^3 p \ \frac{p^2}{3} F(p^2) \ ,
\end{equation}
where $k$ is any of the components $k=1,2,3$.

} $p_i^2 \rightarrow \frac{p^2}{3}$, and so
\begin{equation}
 \partial_{\tau} \left( C^{-4} L_{\tau \tau}(u_{\pmb{p};j},u_{\pmb{p;j}}) \right) + 5 C^{-5} \dot{C} L_{\tau \tau} (u_{\pmb{p};j},u_{\pmb{p};j}) + 3 C^{-5} \dot{C} L_{i i} (u_{\pmb{p},j},u_{\pmb{p};j})= 0 \ .
\end{equation}

Likewise it is easy to show that
\begin{equation}
 \partial_{\tau} \left( C^{-4} L_{\tau \tau}(u_{\pmb{p};j},u^*_{-\pmb{p;j}}) \right) + 5 C^{-5} \dot{C} L_{\tau \tau} (u_{\pmb{p};j},u^*_{-\pmb{p};j}) + 3 C^{-5} \dot{C} L_{i i} (u_{\pmb{p},j},u^*_{-\pmb{p};j}) = \frac{C^{-7}\dot{C}}{2(2\pi)^3} \left[ (\chi^*_{p;j})^2 \left(6p_i^2 - 2p^2 \right) \right] \equiv 0 \ .
\end{equation}
This proves the statement for $\nu = \tau$.

\end{itemize}

\section{Energy-momentum tensor in the late time approximation}\label{EnMom}

In this appendix we show the late time form for $\mathbb{T}_{\mu \nu}^{(MIX}$. It is convenient to split the auxiliary tensor components in its mass and kinetic parts. We define (see Eqs. \eqref{AuxTensC1} to \eqref{AuxTensC4})
\begin{eqnarray*}
 && L_{\tau \tau}^{(MASS)} (u_{\pmb{p};j}, u_{\pmb{p};j}) = \frac{m_j^2}{2 (2 \pi)^3} |\chi_{p;j}|^2 \ ; \ \ \ L_{\tau \tau}^{(MASS)} (u_{\pmb{p};j}, u^*_{-\pmb{p};j}) = \frac{m_j^2}{2 (2 \pi)^3} (\chi^*_{p;j})^2 \ ; \\
 && L_{k k}^{(MASS)} (u_{\pmb{p};j}, u_{\pmb{p};j}) = - L_{\tau \tau}^{(MASS)} (u_{\pmb{p};j}, u_{\pmb{p};j}) \ ; \ \ \ L_{k k}^{(MASS)} (u_{\pmb{p};j}, u^*_{-\pmb{p};j}) = - L_{\tau \tau}^{(MASS)} (u_{\pmb{p};j}, u^*_{-\pmb{p};j})
\end{eqnarray*}
and the ``kinetic'' parts as $L^{(KIN)}_{\mu \mu} (A, B) = L_{\mu \mu} (A, B) - L^{(MASS)}_{\mu \mu} (A,B)$, for $\mu = \tau, k$ and $A,B = u_{\pmb{p};j},u^*_{-\pmb{p};j}$. Letting $(a) = (MASS), (KIN)$, we also set

\begin{eqnarray}
 \nonumber \mathbb{T}_{\mu \mu}^{(a)} &=& \sin^2 \theta \int d^3 p \bigg \lbrace 2 |\Xi_p (\tau_0)|^2 \sum_{j=1,2} L^{(a)}_{\mu \mu} (u_{\pmb{p};j}, u_{\pmb{p};j}) + \left[ \Xi_p^* (\tau_0) \Lambda_p (\tau_0) L^{(a)}_{\mu \mu} (u_{\pmb{p};1},u^*_{-\pmb{p};1}) + c.c. \right] \\
 &-& \left[\Xi_p (\tau_0), \Lambda_p (\tau_0) L^{(a)*}_{\mu \mu} (u_{\pmb{p};2}, u^*_{-\pmb{p};2}) + c.c. \right] \bigg \rbrace \ .
\end{eqnarray}
Evidently
\begin{equation}
 \mathbb{T}^{(MIX)}_{\mu \mu} = \mathbb{T}^{(MASS)}_{\mu \mu} + \mathbb{T}^{(KIN)}_{\mu \mu}
\end{equation}
for $ \mu = \tau, k$. Of course no sum is involved over the repeated index. At the lowest order in $\tau$ we have
\begin{eqnarray}\label{MassIntegral}
 \nonumber &&\mathbb{T}_{\tau \tau}^{(MASS)} (\tau_0, \tau) = - \mathbb{T}_{kk}^{(MASS)} (\tau_0, \tau) = \frac{\sin^2 \theta}{(2\pi)^3} \int d^3 p \Bigg \lbrace \frac{m_1^2 \tau \coth{\pi |\nu_2|}}{8|\nu_1|^2|\nu_2|}\Bigg[ (|\nu_1|^2 + |\nu_2|^2) + \\
 \nonumber &&\frac{(|\nu_1|^2 - |\nu_2|^2)\left(\sinh^2 (\pi |\nu_1|) + \cosh^2 (\pi |\nu_1|) \right)}{4 \sinh^2 (\pi |\nu_1|)}\left( \left(\frac{\tau}{\tau_0}\right)^{2 \nu_1} + \left(\frac{\tau_0}{\tau}\right)^{2 \nu_1} \right) \Bigg] + \frac{m_2^2 \tau \coth{\pi |\nu_1|}}{8|\nu_1||\nu_2|^2}\Bigg[(|\nu_1|^2 + |\nu_2|^2) + \\
 \nonumber &&\frac{(|\nu_2|^2 - |\nu_1|^2)\left(\sinh^2 (\pi |\nu_2|) + \cosh^2 (\pi |\nu_2|) \right)}{4 \sinh^2 (\pi |\nu_2|)}\left( \left(\frac{\tau}{\tau_0}\right)^{2 \nu_2} + \left(\frac{\tau_0}{\tau}\right)^{2 \nu_2} \right) \Bigg] -\sum_j \frac{m_j^2 \tau \coth (\pi |\nu_j|)}{4 |\nu_j|} + \\
 && \nonumber \Bigg \lbrace \left(\frac{p}{2} \right)^{2(\nu_1-\nu_2)} \frac{\pi ^2}{8 \sinh^2(\pi |\nu_1|) \sinh^2 (\pi|\nu_2|) \Gamma^2 (1 + \nu_1) \Gamma^2 (1-\nu_2) } \Bigg[ (-\tau_0)^{2(\nu_1 - \nu_2)} (\nu_1 + \nu_2)^2 \Bigg(  \frac{m_1^2 \tau (\coth (\pi |\nu_1|)-1)}{4 |\nu_j|} \Bigg) \\
 \nonumber && + (-\tau_0)^{-2\nu_2}(-\tau)^{2 \nu_1} (|\nu_1|^2 - |\nu_2|^2)\left(\frac{m_1^2 \tau \coth (\pi |\nu_1|)}{4 |\nu_1|} - \frac{m_1^2 \tau (1 + e^{\pi |\nu_1|})}{8 |\nu_1| \sinh (\pi |\nu_1|)} \right) + (-\tau_0)^{2\nu_1}(-\tau)^{-2 \nu_2} (|\nu_2|^2 - |\nu_1|^2) \times \\
 \nonumber && \left(\frac{m_2^2 \tau \coth (\pi |\nu_2|)}{4 |\nu_2|} - \frac{m_2^2 \tau (1 + e^{\pi |\nu_2|})}{8 |\nu_2| \sinh (\pi |\nu_2|)} \right) \Bigg] + c.c. \Bigg \rbrace + \Bigg \lbrace \left(\frac{p}{2} \right)^{2(\nu_1 + \nu_2)}\frac{\pi ^2 (-\tau_0)^{2\nu_2}(-\tau)^{2\nu_1}(\nu_2 - \nu_1)^2}{8 \sinh^2(\pi |\nu_1|) \sinh^2 (\pi|\nu_2|) \Gamma^2 (1 + \nu_1) \Gamma^2 (1+\nu_2)} \times \\
 \nonumber && \Bigg[\frac{m_1^2 \tau \coth (\pi |\nu_1|))}{4 |\nu_1|} - \frac{m_1^2 \tau (1+ e^{\pi |\nu_1|})}{8 |\nu_1| \sinh (\pi |\nu_1|)} -\frac{m_2^2 \tau \coth (\pi |\nu_2|))}{4 |\nu_2|} +  \frac{m_2^2 \tau (1+ e^{\pi |\nu_2|})}{8 |\nu_2| \sinh (\pi |\nu_2|)} \Bigg] + c.c. \Bigg \rbrace + \\
 && \nonumber \Bigg \lbrace \left(\frac{p}{2} \right)^{2 \nu_1} \frac{\pi}{4 |\nu_2| \sinh^2 (\pi |\nu_1|) \Gamma^2 (1+ \nu_1)} \Bigg[(- \tau_0)^{2\nu_1} (|\nu_1|^2 - |\nu_2|^2)\left(\frac{m_2^2 \tau}{4|\nu_2|}\right) + (-\tau_0)^{2(\nu_1-\nu_2)}(-\tau)^{2\nu_2} (\nu_1 + \nu_2)^2 \times \\
 \nonumber && \left( \frac{m_2^2 \tau}{16 |\nu_2| \sinh^2 (\pi |\nu_2|)}\right) \left(e^{2 \pi |\nu_2|} + e^{- \pi |\nu_2|} -1  \right) + (-\tau_0)^{2(\nu_1 + \nu_2)} (-\tau)^{-2\nu_2} (\nu_2 - \nu_1)^2 \left( \frac{m_2^2 \tau}{16 |\nu_2| \sinh^2 (\pi |\nu_2|)}\right) \times \\
 \nonumber && \left(e^{2 \pi |\nu_2|} + e^{- \pi |\nu_2|} -1  \right)\Bigg] + c.c. \Bigg \rbrace + \Bigg \lbrace \left(\frac{p}{2} \right)^{2 \nu_2}\frac{\pi}{4 |\nu_1| \sinh^2 (\pi |\nu_2|) \Gamma^2 (1+ \nu_2)} \Bigg[(- \tau_0)^{2\nu_2} (|\nu_2|^2 - |\nu_1|^2)\left(\frac{m_1^2 \tau}{4|\nu_1|}\right) + \\
 \nonumber && (-\tau_0)^{2(\nu_2-\nu_1)}(-\tau)^{2\nu_1} (\nu_1 + \nu_2)^2\left( \frac{m_1^2 \tau}{16 |\nu_1| \sinh^2 (\pi |\nu_1|)}\right) \left(e^{2 \pi |\nu_1|} + e^{- \pi |\nu_1|} -1  \right) + (-\tau_0)^{2(\nu_1 + \nu_2)} (-\tau)^{-2\nu_1} (\nu_2 - \nu_1)^2 \times \\
 && \left( \frac{m_1^2 \tau}{16 |\nu_1| \sinh^2 (\pi |\nu_1|)}\right) \left(e^{2 \pi |\nu_1|} + e^{- \pi |\nu_1|} -1  \right) \Bigg] + c.c. \Bigg \rbrace \Bigg \rbrace \ ,
\end{eqnarray}
\begin{eqnarray}\label{EnergyIntegral}
 \nonumber && \mathbb{T}_{\tau \tau}^{(KIN)} = \frac{\sin^2 \theta H_0^2 \tau}{(2\pi)^3} \int d^3 p \Bigg \lbrace \left(\frac{\coth (\pi |\nu_1|) \coth (\pi |\nu_2|) (|\nu_1|^2 + |\nu_2|^2)}{2 |\nu_1| |\nu_2|} -1 \right) \sum_j \frac{\coth (\pi |\nu_j|)}{|\nu_j|} \left(\frac{1}{2}-\frac{m_j^2}{4 H_0^2} \right) + \\
 \nonumber && \frac{\coth(\pi |\nu_2|)(|\nu_1|^2 + |\nu_2|^2)}{4 |\nu_1|^2 |\nu_2| \sinh^2 (\pi |\nu_1|)} \left( \frac{m_1^2}{4H_0^2} - \frac{1}{2}\right) + \frac{\coth(\pi |\nu_1|)(|\nu_1|^2 + |\nu_2|^2)}{4 |\nu_1| |\nu_2|^2 \sinh^2 (\pi |\nu_2|)} \left( \frac{m_2^2}{4H_0^2} - \frac{1}{2}\right) + \Bigg[ \left(\frac{\tau}{\tau_0} \right)^{2 \nu_2} \frac{\coth(\pi |\nu_1|)(|\nu_2|^2 -|\nu_1|^2)}{8 |\nu_1||\nu_2|^2 \sinh^2 (\pi |\nu_2|)} \times \\
 \nonumber && \left(\left(\frac{5}{8} - \frac{m_2^2}{4H_0^2} - \frac{\nu_2}{4} \right) + \cosh(2\pi |\nu_2|)\left( \frac{-5}{8} + \frac{m_2^2}{4H_0^2} -\frac{\nu_2}{4}\right) \right) + c.c. \Bigg] + \Bigg[ \left(\frac{\tau}{\tau_0} \right)^{2 \nu_1} \frac{\coth(\pi |\nu_2|)(|\nu_1|^2-|\nu_2|^2)}{8 |\nu_1|^2|\nu_2| \sinh^2 (\pi |\nu_1|)} \times \\
 \nonumber && \left( \left(\frac{5}{8}- \frac{m_1^2}{4H_0^2} - \frac{\nu_1}{4} \right) + \cosh( 2\pi |\nu_1| ) \left( \frac{-5}{8} + \frac{m_1^2}{4H_0^2}- \frac{\nu_1}{4}\right) \right) + c.c. \Bigg] + \\
 \nonumber && \Bigg[\left(\frac{p}{2} \right)^{2(\nu_1 - \nu_2)} \frac{\pi^2}{16 \sinh^2 ( \pi |\nu_1|) \sinh^2 (\pi |\nu_2|)\Gamma^2(1+ \nu_1) \Gamma^2 (1-\nu_2)} \Bigg( (- \tau_0)^{2\nu_1} (-\tau)^{-2\nu_2} (|\nu_1|^2 - |\nu_2|^2)\coth (\pi |\nu_2|) \nu_2 - \\
 \nonumber && (-\tau_0)^{- 2\nu_2}(- \tau)^{2 \nu_1} (|\nu_2|^2 - |\nu_1|^2)\coth(\pi |\nu_1|)\nu_1 \Bigg) + c.c. \Bigg] + \\
 \nonumber && \Bigg[ \left(\frac{p}{2} \right)^{2(\nu_1 + \nu_2)} \frac{\pi^2}{16 \sinh^2 (\pi |\nu_1|) \sinh^2 (\pi |\nu_2| \Gamma^2 (1+\nu_1) \Gamma^2 (1+ \nu_2)} \Bigg( - (-\tau_0)^{2 \nu_1} (-\tau)^{2 \nu_2} (|\nu_1|^2 - |\nu_2|^2) \coth(\pi |\nu_2|) \nu_2 - \\
 \nonumber && (-\tau_0)^{2\nu_2} (-\tau)^{2 \nu_1}(|\nu_2|^2 - |\nu_1|^2)\coth(\pi |\nu_1|)\nu_1 \Bigg) + c.c. \Bigg] + \\
 \nonumber && \Bigg[ \left(\frac{p}{2} \right)^{2\nu_1} \frac{\pi}{8 |\nu_2| \sinh^2 (\pi |\nu_1|) \sinh^2 (\pi |\nu_2|)\Gamma^2 (1+\nu_1)} \Bigg( 2 (\tau_0)^{2\nu_1} \frac{(|\nu_1|^2 - |\nu_2|^2)}{|\nu_2|}\left( \frac{1}{2} - \frac{m_2^2}{4H_0^2} \right)  \sinh^2 (\pi |\nu_2|) + \\
 \nonumber && (-\tau_0)^{2 (\nu_1 - \nu_2)} (-\tau)^{2 \nu_2} \frac{(\nu_1 + \nu_2)^2}{2 |\nu_2|}\left( (1-2 \cosh(2\pi |\nu_2|))\left(\frac{5}{8}-\frac{m_2^2}{4H_0^2} \right) - \left( 1 + 2 \cosh(2\pi |\nu_2|)\right)\frac{\nu_2}{4} \right) + \\
 \nonumber &&  (-\tau_0)^{2(\nu_1 + \nu_2)} (-\tau)^{-2\nu_2} \frac{(\nu_1 - \nu_2)^2}{2|\nu_2|}\left( (1-2 \cosh(2\pi |\nu_2|))\left(\frac{5}{8}-\frac{m_2^2}{4H_0^2} \right) + \left( 1 + 2 \cosh(2\pi |\nu_2|)\right)\frac{\nu_2}{4} \right) - \\
 \nonumber && (-\tau)^{2\nu_2} \frac{4}{|\nu_1|} \left(|\nu_1||\nu_2| \sinh^2 (\pi |\nu_2|) \left(\frac{5}{8} - \frac{m_1^2}{4H_0^2} - \frac{\nu_1}{4} \right) + \frac{\nu_1 (|\nu_1|^2 + |\nu_2|^2)}{4} \coth( \pi |\nu_1|) \cosh (\pi |\nu_2|) \sinh (\pi |\nu_2|)\right) \Bigg) + c.c. \Bigg] + \\
  \nonumber && \Bigg[ \left(\frac{p}{2} \right)^{2\nu_2} \frac{\pi}{8 |\nu_1| \sinh^2 (\pi |\nu_1|) \sinh^2 (\pi |\nu_2|)\Gamma^2 (1+\nu_2)} \Bigg( 2 (\tau_0)^{2\nu_2} \frac{(|\nu_2|^2 - |\nu_1|^2)}{|\nu_1|}\left( \frac{1}{2} - \frac{m_1^2}{4H_0^2} \right)  \sinh^2 (\pi |\nu_1|) + \\
 \nonumber && (-\tau_0)^{2 (\nu_2 - \nu_1)} (-\tau)^{2 \nu_1} \frac{(\nu_1 + \nu_2)^2}{2 |\nu_1|}\left( (1-2 \cosh(2\pi |\nu_1|))\left(\frac{5}{8}-\frac{m_1^2}{4H_0^2} \right) - \left( 1 + 2 \cosh(2\pi |\nu_1|)\right)\frac{\nu_1}{4} \right) + \\
 \nonumber &&  (-\tau_0)^{2(\nu_1 + \nu_2)} (-\tau)^{-2\nu_1} \frac{(\nu_1 - \nu_2)^2}{2|\nu_1|}\left( (1-2 \cosh(2\pi |\nu_1|))\left(\frac{5}{8}-\frac{m_1^2}{4H_0^2} \right) + \left( 1 + 2 \cosh(2\pi |\nu_1|)\right)\frac{\nu_1}{4} \right) - \\
 \nonumber && (-\tau)^{2\nu_1} \frac{4}{|\nu_2|} \left(|\nu_1||\nu_2| \sinh^2 (\pi |\nu_1|) \left(\frac{5}{8} - \frac{m_2^2}{4H_0^2} - \frac{\nu_2}{4} \right) + \frac{\nu_2 (|\nu_1|^2 + |\nu_2|^2)}{4} \coth( \pi |\nu_2|) \cosh (\pi |\nu_1|) \sinh (\pi |\nu_1|)\right) \Bigg) + c.c. \Bigg] \Bigg \rbrace \ ,\\
 &&
\end{eqnarray}
\begin{eqnarray}\label{PressureIntegral}
\nonumber && \mathbb{T}_{kk}^{(KIN)} = \frac{\sin^2 \theta}{(2\pi)^3} \int d^3 p \Bigg \lbrace \sum_j \frac{m_j^2 \tau \coth (\pi |\nu_j|)}{4|\nu_j|} + \Bigg[\left(\frac{\tau}{\tau_0} \right)^{2\nu_2} \frac{H_0^2 \tau \coth(\pi |\nu_1|) (|\nu_2|^2 - |\nu_1|^2)}{8 |\nu_1| |\nu_2|^2 \sinh^2 (\pi |\nu_2|)}  \left(\frac{9}{8} - \frac{m_2^2}{4H_0^2} + \frac{3\nu_2}{4} \right) \times \\
\nonumber && \left(1- \cosh(2\pi |\nu_2|) \right) + c.c. \Bigg] + \Bigg[\left(\frac{\tau}{\tau_0} \right)^{2\nu_1} \frac{H_0^2 \tau \coth(\pi |\nu_2|) (|\nu_1|^2 - |\nu_2|^2)}{8 |\nu_2| |\nu_1|^2 \sinh^2 (\pi |\nu_1|)}  \left(\frac{9}{8} - \frac{m_1^2}{4H_0^2} + \frac{3\nu_1}{4} \right)  \left(1- \cosh(2\pi |\nu_1|) \right) + c.c. \Bigg] +\\
\nonumber && \Bigg[ \left(\frac{p}{2} \right)^{2\nu_1} \frac{\pi}{4 |\nu_2| \sinh^2(\pi |\nu_1|) \Gamma^2 (1+\nu_1)} \Bigg((-\tau_0)^{2\nu_1} \frac{m_2^2 \tau (|\nu_2|^2 - |\nu_1|^2)}{4|\nu_2|} + (-\tau_0)^{2(\nu_1 - \nu_2))} (-\tau)^{2\nu_2} \frac{(\nu_1 + \nu_2)^2 H_0^2 \tau)}{4|\nu_2| \sinh^2 (\pi |\nu_2|)} \times \\
\nonumber && \left(\frac{9}{8} - \frac{m_2^2}{4H_0^2} + \frac{3\nu_2}{4} \right)\left(1- \cosh (2 \pi |\nu_2|) \right) + (-\tau_0)^{2(\nu_1 + \nu_2)} (-\tau)^{-2\nu_2} \frac{(\nu_2 - \nu_1)^2 H_0^2 \tau}{4|\nu_2|\sinh^2 (\pi |\nu_2|))}\left(\frac{9}{8}-\frac{m_2^2}{4H_0^2}-\frac{3\nu_2}{4} \right) \times \\
\nonumber && \left(1 - \cosh(2\pi |\nu_2|) \right)-2(-\tau)^{2\nu_1} H_0^2 \tau |\nu_2| \left(\frac{9}{8}- \frac{m_1^2}{4H_0^2} + \frac{3\nu_1}{4} \right) \Bigg) + c.c \Bigg] + \\
\nonumber && \Bigg[ \left(\frac{p}{2} \right)^{2\nu_2} \frac{\pi}{4 |\nu_1| \sinh^2(\pi |\nu_2|) \Gamma^2 (1+\nu_2)} \Bigg((-\tau_0)^{2\nu_2} \frac{m_1^2 \tau (|\nu_1|^2 - |\nu_2|^2)}{4|\nu_1|} + (-\tau_0)^{2(\nu_2 - \nu_1))} (-\tau)^{2\nu_1} \frac{(\nu_1 + \nu_2)^2 H_0^2 \tau)}{4|\nu_1| \sinh^2 (\pi |\nu_1|)} \times \\
\nonumber && \left(\frac{9}{8} - \frac{m_1^2}{4H_0^2} + \frac{3\nu_1}{4} \right)\left(1- \cosh (2 \pi |\nu_1|) \right) + (-\tau_0)^{2(\nu_1 + \nu_2)} (-\tau)^{-2\nu_1} \frac{(\nu_2 - \nu_1)^2 H_0^2 \tau}{4|\nu_1|\sinh^2 (\pi |\nu_1|))}\left(\frac{9}{8}-\frac{m_1^2}{4H_0^2}-\frac{3\nu_1}{4} \right) \times \\
 && \left(1 - \cosh(2\pi |\nu_1|) \right)-2(-\tau)^{2\nu_2} H_0^2 \tau |\nu_1| \left(\frac{9}{8}- \frac{m_2^2}{4H_0^2} + \frac{3\nu_2}{4} \right) \Bigg) + c.c \Bigg] \Bigg \rbrace \ .
\end{eqnarray}

\end{document}